\documentclass[aps,prb,preprint,superscriptaddress,showpacs,floatfix]{revtex4-2}
\usepackage{amsmath,amssymb,amsfonts}
\usepackage{graphicx}
\usepackage{bm}
\usepackage{siunitx}

\begin{document}
\title{Real-space manifestation of ferroic multipoles in altermagnetic MnF$_2$}

\author{Iurii Kibalin}
\affiliation{Data Management and Scientific Computing,
European Spallation Source,  ERIC,
2800 Kongens Lyngby, Denmark}

\author{Dalila Bounoua}
\affiliation{Laboratoire L\'eon Brillouin,
CEA, CNRS, 
CEA-Saclay,
91191 Gif-sur-Yvette, France}
\author{Jos\'e A.~Rodriguez Velamaz\'an}
\affiliation{Institut Laue-Langevin, 71 avenue des Martyrs,
38042 Grenoble, France}
\author{Oscar Fabelo}
\affiliation{Institut Laue-Langevin, 71 avenue des Martyrs,
38042 Grenoble, France}
\author{Navid Qureshi}
\affiliation{Institut Laue-Langevin, 71 avenue des Martyrs,
38042 Grenoble, France}
\author{Quentin Faure}
\affiliation{Laboratoire L\'eon Brillouin,
CEA, CNRS, 
CEA-Saclay,
91191 Gif-sur-Yvette, France}
\author{Philippe Bourges}
\affiliation{Laboratoire L\'eon Brillouin,
CEA, CNRS, 
CEA-Saclay,
91191 Gif-sur-Yvette, France}
\author{Victor Bal\'edent}
\affiliation{Laboratoire de Physique des Solides,
CNRS, Universit\'e Paris-Saclay,
91405 Orsay, France}
\affiliation{Laboratoire L\'eon Brillouin,
CEA, CNRS, 
CEA-Saclay,
91191 Gif-sur-Yvette, France}
\author{Jian-Rui Soh}
\affiliation{Quantum Innovation Centre (Q.InC),
 (A*STAR),
 138634, Singapore}
\affiliation{Centre for Quantum Technologies,
National University of Singapore,
 117543, Singapore}
\author{Jeffrey Rau}
\affiliation{Department of Physics,
University of Windsor,
Windsor, Ontario N9B 3P4, Canada}

\author{Paul McClarty}
\affiliation{Laboratoire L\'eon Brillouin,
CEA, CNRS, 
CEA-Saclay,
91191 Gif-sur-Yvette, France}
\author{Arsen Gukasov}
\affiliation{Laboratoire L\'eon Brillouin,
CEA, CNRS, 
CEA-Saclay,
91191 Gif-sur-Yvette, France}


\begin{abstract}
Altermagnets are unconventional spin split magnets arising from the zero spin-orbit coupled limit. They host a magnetic multipolar order parameter yet direct real-space observation of these multipoles has remained elusive. Here we use polarized-neutron diffraction to reconstruct the three-dimensional magnetization density of the prototypical altermagnet MnF$_2$. By exploiting symmetry-selective magnetic reflections, we separate the dominant spherical Mn$^{2+}$ contribution from the much weaker anisotropic Mn magnetization and the covalent spin polarization of the fluorine ligands. The reconstructed spin density reveals a finite fluorine ion moment  together with an anisotropic Mn magnetization consistent with the symmetry-allowed altermagnetic rank-5 magnetic multipole $O_{52}$(magnetic triacontadipole). These results provide direct real-space evidence of ferroic multipolar order in an altermagnet and establish polarized-neutron diffraction as a powerful probe of hidden magnetic multipoles in quantum materials.
\end{abstract}

\maketitle

\section{Introduction}\label{sec:introduction}

Altermagnets are a class of compensated collinear magnets with spin-split electronic bands allowed by their special magneto-crystalline symmetries \cite{naka2019,Smejkal2019,hayami2019,yuan2020,Smejkal2023}. Unlike simple antiferromagnets, the magnetic sublattices of altermagnets are connected by lattice rotation or mirror operations. In the zero spin-orbit coupling limit where altermagnetism is most cleanly defined, the electronic bands have the spin projection as a good quantum number and magnons have a well-defined chirality and both exhibit a spin reversal under the same point group operation that swaps the magnetic sublattices \cite{naka2019,Smejkal2019,hayami2019,Smejkal2023}. The nature of the anisotropic momentum space spin-splitting lends itself to a natural classification into d-wave, g-wave and i-wave systems. The magnetic anisotropy of altermagnets combined with magnetic compensation are very attractive for potential spintronics applications in real materials as spin-orbit coupling may only weakly perturb the principal zero spin-orbit coupling effect \cite{Smejkal2022b}.

 To date, several altermagnetic materials have been characterized as such via transport measurements and spectroscopy \cite{jungwirth2025review} for example in the now canonical examples MnTe \cite{gonzalez,lee2024,liu2024,osumi2024,Hajlaoui} and CrSb \cite{reimers2024,zeng2024,yang2025,ding2024,lu2024crsb}. Here we focus on the magnetic insulator MnF$_2$. Altermagnetism in this material has been characterized through direct visualization of the opposed spin polarizations of the magnonic bands using polarized inelastic neutron scattering \cite{faure2025}. 

The magnetic anisotropy of altermagnets in momentum space is the momentum space realization of some distinctive local order parameter. While the staggered magnetization is a common order parameter in all altermagnets, it typically has the feature that the magnetic sublattice {\it considered alone} has a higher non-altermagnetic symmetry than that of the full magneto-crystalline structure. Therefore, any local order parameter respecting the altermagnetic symmetries must be some kind of multipole \cite{Smejkal2019,hayami2019,hayami2020bottomup,bhowal2024,mcclarty2024,schiff2025}. In practice, one may construct such multipoles directly from clusters of several atomic sites \cite{hayami2019,hayami2020bottomup}. However, first principles calculations on MnF$_2$ revealed a distinctively altermagnetic multipolar contribution in the form of specific ferroic multipoles in the local magnetization density \cite{bhowal2024}. Such ferroic multipoles generalize to all altermagnets inherited from the pristine zero spin-orbit coupled limit \cite{verbeek2024,mcclarty2024,schiff2025}. While all altermagnetic properties could be viewed as being tied to such multipolar order on symmetry grounds, it is of interest to image the multipoles directly.

Building on earlier work on MnF$_2$ \cite{costa1989, nathans1963} in this work, we employ polarized neutron diffraction (PND) to reconstruct the full three-dimensional magnetization density of MnF$_2$ in real space.
This technique is uniquely suited to probing very subtle magnetic features, revealing both the radial extent of the unpaired electrons and deviations from spherical symmetry, as well as covalent spin density transferred to ligand atoms. Using high performance PND diffractometer D3 at the ILL  and newly developed software based on magnetic space-group theory,  maximum entropy method (MEM) and multipole analysis \cite{cryspy}, we quantitatively determine the covalent spin density on the fluorine ligands. We further analyze the anisotropy of the Mn spin density, giving evidence for the  altermagnetic multipolar order in this compound.

Although polarized neutron diffraction  has proven highly effective for ferro- and ferrimagnets, it has rarely been applied to compensated magnets \cite{jeong2020}. In these latter systems, a measurable polarization dependence arises only under the conditions that magnetic atoms carrying opposite spins must be related by a symmetry element of the space group other than a lattice translation or inversion symmetry \cite{Alperin1962,mcclarty2025,schiff2025}.  These are precisely the necessary conditions for altermagnetism making them excellent candidate materials for further investigation using polarized neutrons.

We organize the paper as follows. In the following section (Section~\ref{sec:results}), we briefly introduce the polarized neutron diffraction technique and some of the specifics of the MnF$_2$ experiment.  Section~\ref{subsec1} introduces the symmetry-allowed multipoles in MnF$_2$. We then present results of the diffraction experiment. On the basis of these results we show (Section~\ref{subsec2}) that, in addition to spontaneous moments being present on the magnetic manganese ions, the fluorine moments in MnF$_2$ are also magnetically ordered in such a way that reflects the underlying altermagnetic symmetries. We point out that the fluorine moments make a contribution to a magnetic multipolar order parameter connected to the altermagnetism. Then, in Section~\ref{subsec3} we show results of a maximum entropy fit to obtain the microscopic magnetization density within the unit cell. This then provides (Section~\ref{subsec4}) evidence for the presence of asphericities of the magnetization density around the manganese ions that contribute to altermagnetic multipolar order parameters. We comment on plausible implications of these magnetization densities for the altermagnetic exchange pathways in the supplementary material \ref{sec:supplementary}.

\section{Methods}\label{sec:methods}

Polarized neutron diffraction was carried out on a high quality single crystal of MnF$_2$ on the D3 spectrometer at the ILL facility. The measured scattering intensity has three contributions of interest $I^{\pm} = I_{\rm n}+I_{\rm m}\pm I_{\rm nm}$ the first term originating from the nuclear structure and the second from the magnetic structure. These two terms are time reversal even. The third term is the nuclear-magnetic interference term. Unlike the first two terms, it depends on the incident neutron polarization reversing sign when the polarization is flipped or when the intrinsic moments are reversed. For the experimental refinement it is useful to introduce the flipping ratio $R=I^{+}/I^{-}$ and asymmetry parameter $As = (I^{+}-I^{-})/(I^{+}+I^{-}) = 2I_{\rm nm}/(I_{\rm n}+I_{\rm m})$ $-$ the latter providing direct insight into the polarization dependent interference term relative to the remaining contributions to the intensity.

In $\mathrm{MnF_2}$, it is useful to separate out contributions from the \textit{even} ($h+k+l=2n$) and \textit{odd} reflections ($h+k+l=2n+1$). For the nuclear structure on its own, the manganese sublattice contributes only to the even reflections while fluorine nuclei contribute to both.  Turning to the magnetic signal, the simple antiferromagnetic structure on the Mn sites (the spherical Mn contribution) is the dominant contribution and appears
only in the odd reflections but in both $I_{\rm m}$ and $I_{\rm nm}$. The magnetic contribution to the even peaks belongs to any covalency on the fluorine sites and any asphericity on the manganese sites. Therefore the even peaks are of particular interest in connection to altermagnetism.

The  magnetic structure of MnF$_2$ supports two 180$^\circ$ antiferromagnetic domains, distinguished by the sign of the N\'eel vector  $\mathbf{N}$, and denoted $\mathbf{N}_{\pm}$. In order to make use of the neutron polarization it was necessary to achieve a majority single-domain state. Field-cooling at 1~T  yielded a nearly single-domain state (98\% $\mathbf{N}_{-}$), with the N\'eel vector antiparallel to the applied field. A separate field cooling experiment down to 6~T inverted the domain population relative to the first experiment, yielding approximately 86\% of the $\mathbf{N}_{+}$ domain consistent with earlier observations~\cite{Felcher1996}. For further details of the domain selection see the supplementary section.

\section{Results and discussion}\label{sec:results}

\subsection{Characterization of altermagnets through multipolar order}\label{subsec1}

Altermagnets are collinear compensated magnets with a net staggered magnetization $\mathbf{N}=\mathbf{M}_A-\mathbf{M}_B$ where $\mathbf{M}_I$ is the magnetization on sublattice $I=A,B$. Taking into account the full magneto-crystalline symmetries and assuming that the magnetic order does not break lattice translation symmetry, altermagnetism is characterized by $\mathbf{N}$ transforming as a 1D non-trivial irreducible representation of the point group of the lattice. Looking for order parameters at zero spin-orbit coupling with identical transformation properties one finds in all altermagnets a multipolar order parameter of the form
\begin{equation}
\mathbf{O}_{m_1  \ldots m_p} = \int d^3\mathbf{r} [r_{m_1} \ldots r_{m_p}] \mathbf{m}(\mathbf{r})
\end{equation}
where $[\ldots]$ is a real space multipole and $\mathbf{m}(\mathbf{r})$ is the rotationally symmetric local magnetization density. As this transforms like the N\'{e}el order parameter (once all nonmagnetic sites are included) \cite{mcclarty2024}, cooling through $T_N$ also leads to a nonzero expectation value for this multipole. 

In other words, each local moment is associated with some generally aspherical local magnetization density that can be expanded in time odd multipoles. Among allowed multipoles, there is one ferroic multipole (preserving sign from one magnetic sublattice to another), to leading order, whose intrinsic transformation properties directly capture the  magneto-crystalline symmetries.  

For example, MnF$_2$, (characterized by magnetic space group $P4_2'$/mnm$'$) contains two magnetic sublattices with equal and opposite moments related by a four-fold rotation around the axis of the moment, time reversal and $\frac{1}{2}\frac{1}{2}\frac{1}{2} $  translation. In the presence of  magnetic anisotropies that pin the local moments along the $c$ axis the Landau theory leads to an $xy m_z$ multipole to leading order in the multipole expansion \cite{bhowal2024} descended from the $xy \mathbf{m}$ multipole coming from the spin-orbit free limit. This multipole respects the altermagnetic symmetries. Combining the real space and magnetic components gives a single magnetic multipole with $O_{32-}$ character. This notation for the multipoles generalizes to $O_{lm}$. For this instance, $l=3$ means that the object is an octupole while $m=2$ specifies the components $(x\pm iy)^2m_z$. The multipole is time reversal odd and is therefore magnetic. Later in this work, we shall see that a symmetry-allowed multipole of higher order $-$ $xy z^2 m_z$ $-$ is important in MnF$_2$. The existence of this multipole, $O_{52}$, like that of its lower order counterpart $O_{32}$, captures altermagnetic symmetries in a local order parameter. Below we discuss other symmetry-allowed multipoles in MnF$_2$.

\subsection{Polarized neutron diffraction}\label{subsec2}

An initial refinement of the flipping ratios $R$ was performed within the \textsc{CRYSPY} software~\cite{cryspy} on the 1~T dataset assuming that the magnetization resides
solely on the Mn sites and follows the spherical Mn$^{2+}$ free-ion form
factor~\cite{FF}. The \textit{odd} reflections
exhibit very large flipping ratios and are therefore sensitive to extinction and
multiple scattering. These effects were minimized by using a thin plate-shaped
sample and by applying extinction corrections. Extinction corrections have a negligible effect on the even reflections.  This model yielded a Mn moment of
$m_{\mathrm{Mn}} = -5.18(4)\,\mu_B$ but produced a poor agreement factor
($\chi^2 = 63.4$), with the largest discrepancies occurring for the
\textit{even} reflections. These reflections are particularly sensitive to
magnetization on the fluorine sites, indicating that a purely Mn-centered model
is insufficient as concluded in Refs.~\cite{costa1989,nathans1963,Felcher1996}.

\begin{figure}[t]
\centering
\includegraphics[width=0.4\columnwidth]{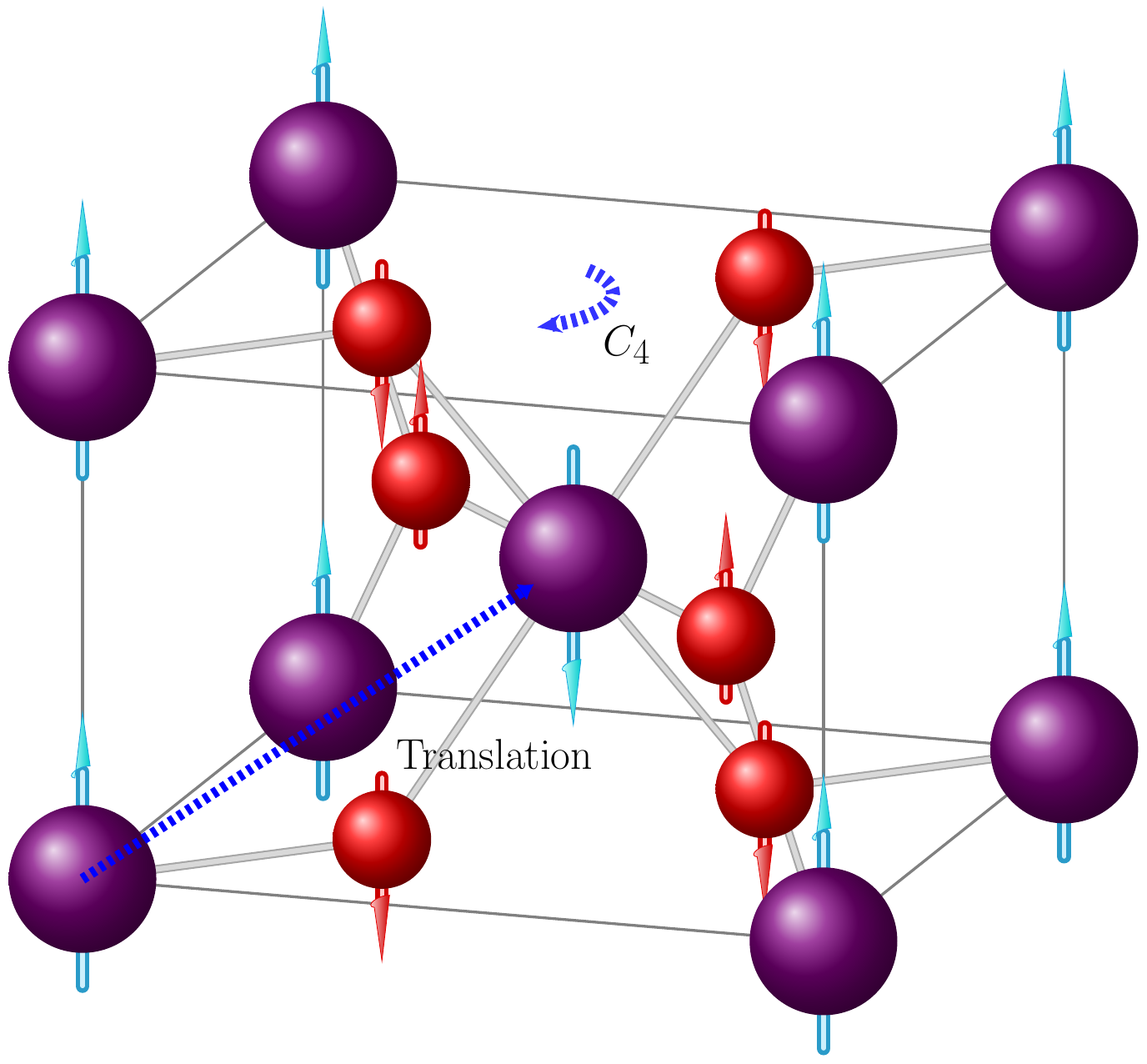}
\caption{
\textbf{Magnetic structure of MnF$_2$.}
 Antiferromagnetic arrangement of Mn moments  at 1~T, including the small but finite ligand spin polarization
on the fluorine sites. The four-fold screw axis reverses the ligand moment, a hallmark of
altermagnetic symmetry. The moments on the fluorine ions have been multiplied by a factor $100$ to make them more clearly visible.
}
\label{fig:MnF2structure}
\end{figure}

Introducing a collinear moment on the fluorine atoms can be done without breaking the
$P4_{2'}/\mathrm{mnm}'$ magnetic symmetry. The symmetry fixes the fluorine
moments to the pattern characteristic of
altermagnetic ordering: the four-fold screw axis reverses the ligand moment and 
the inversion symmetry fixes the moments of other two  fluorine sites. 
Allowing for a finite fluorine moment dramatically improved the refinement,
reducing the agreement factor to $\chi^2 = 5.58$ and yielding
\begin{equation}
m_{\mathrm{Mn}} = -5.18(2)\,\mu_B,\qquad
m_{\mathrm{F}} = 0.028(1)\,\mu_B.
\end{equation}
This refinement reveals a small but clearly resolved fluorine moment antiparallel to the nearest neighbor Mn
moment giving quantitative confirmation of the result in Ref.~\cite{costa1989} and also supported by recent first principles calculations \cite{yuan2020}. The refined magnetic structure for the $\mathbf{N}_{-}$ domain at 1~T is
shown in Fig.~\ref{fig:MnF2structure}. This result for the fluorine moment is consistent with the bound given in Refs.~\cite{costa1989, nathans1963}.

The 6~T dataset was analyzed in the same manner, with the domain population
included as a refinement parameter. The refinement confirmed that after crossing
the spin-flop transition the domain balance was inverted, with $86(2)\%$ of the
sample in the $\mathbf{N}_{+}$ domain. Neglecting the fluorine moment again led
to a poor fit ($\chi^2 = 34.5$), whereas including it produced an excellent
agreement:
\begin{equation}
m_{\mathrm{Mn}} = 5.14(2)\,\mu_B,\qquad
m_{\mathrm{F}} = -0.028(2)\,\mu_B,
\end{equation}
with $\chi^2 = 4.17$ and the expected sign reversal.

The inversion of the fluorine moment under the four-fold screw axis is a direct manifestation of altermagnetic symmetry. Moreover, the ensemble of the four fluorine magnetic moments in the unit cell forming an octahedron around Mn ion can be viewed as a cluster ($\mathbf{k}=0$) magnetic octupole \cite{suzuki2017,hayami2024multipole} expressed as
\begin{equation}
O_{32}(\mathbf{k}=0) = \sum_{i=1}^{4} x_{\mathrm{i,F}}\, y_{\mathrm{i,F}}\, m_{\mathrm{i,F}}.
\end{equation}
Such octupolar order is expected to induce an asphericity of the Mn spin density, which in turn may contribute to the odd magnetic reflections. To quantify the Mn spin-density asphericity, a more powerful approach is required; we therefore turn to a maximum-entropy-method (MEM) reconstruction. For further details see the supplementary material.

\subsection{Maximum Entropy Method}\label{subsec3}
For collinear magnetic structures with real nuclear structure factors $F_N$, the
solution of equation for flipping ratios given in Method section  yields the magnetic structure factors
$M(\mathbf{k})$, which are the Fourier coefficients of the
longitudinal magnetization density in the unit cell,
\begin{equation}
M(\mathbf{k}) =
\int_{\mathrm{cell}} m(\mathbf{r})\, e^{i\mathbf{k}\cdot\mathbf{r}}\, d^3\mathbf{r}.
\end{equation}
A classical Fourier inversion of these coefficients, or a more sophisticated
reconstruction using the Maximum Entropy Method (MEM), provides direct
real-space information on the magnetization density. MEM is known to yield
substantially cleaner real-space densities than conventional Fourier syntheses
by suppressing noise and truncation artifacts.

For this work, we developed a MEM procedure adapted to antiferromagnets. It is similar in spirit to the two-channel approach developed in Ref.~\cite{papoular}, but
incorporates the full magnetic space group and allows constrained magnetic
moments in the asymmetric unit~\cite{karmeshu2003}. The refined Mn and F moments,
$\mu_{\mathrm{Mn}} = -5.18\,\mu_B$ and $\mu_{\mathrm{F}} = 0.028\,\mu_B$, were
assigned to two independent positive density channels, respectively.

\begin{figure*}[t]
    \centering
    \includegraphics[width=\textwidth]{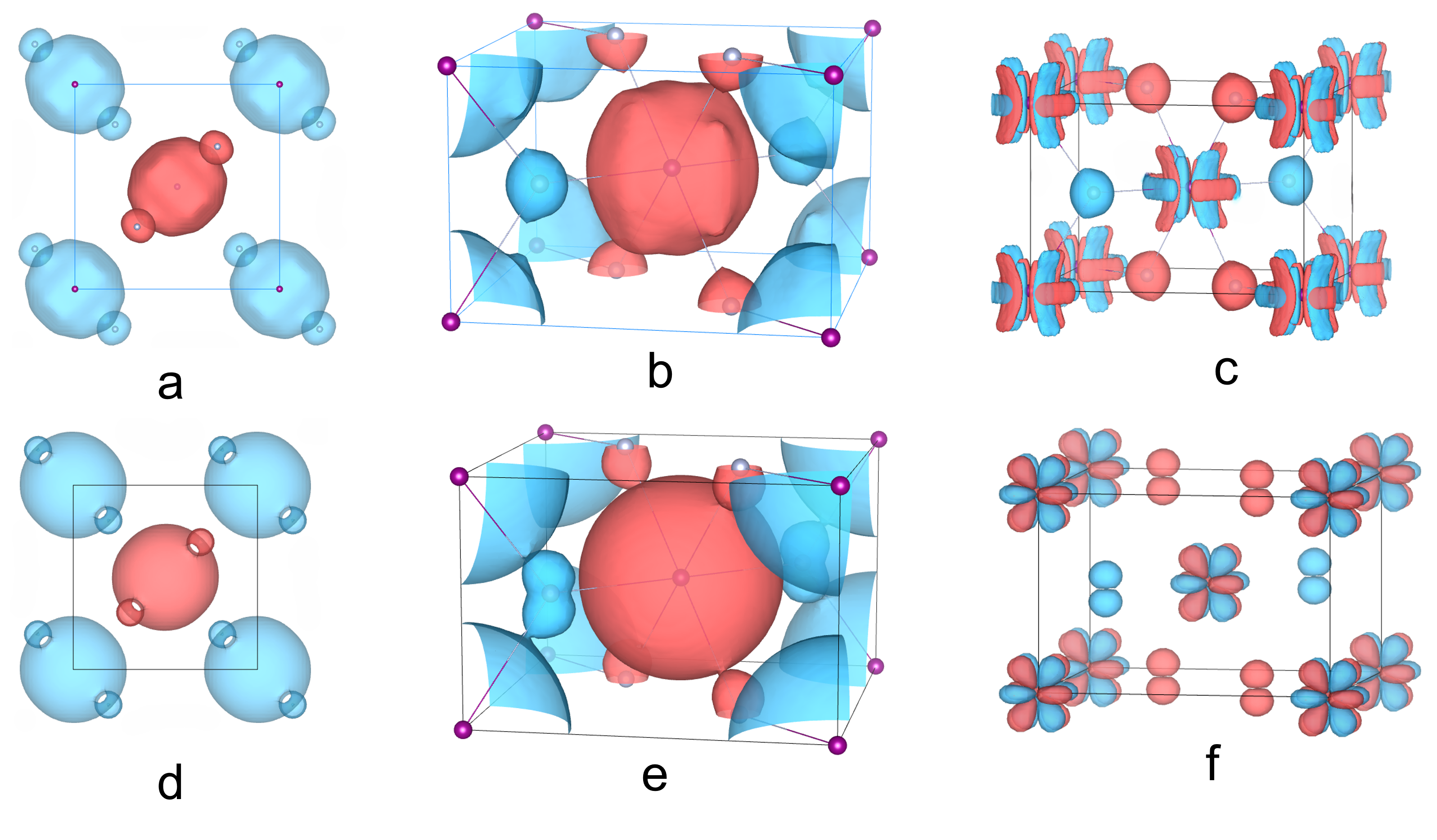}
    \caption{Spin density reconstructed by MEM projected on  the basal plane (a) and in three dimensions (b) at 2~K and 1~T.  Spin density deformation map of the 1~T dataset obtained by subtraction of Mn spherical contribution, demonstrating the multipolar character of the Mn spin distribution (c). Model spin density  obtained by multipole refinement  of the 1T dataset projected on  the basal plane described in the main text (d) and in three dimensions (e) showing the fluorine density exhibiting a $p_{z}$-like character. Model spin density after subtracting the dominant spherical Mn contribution revealing ferroic order of the $O_{52}$ rank-5 magnetic multipole (f).
    Isosurface level of
$0.02\,\mu_B/\text{\AA}^3$ is shown in all figures.}
    \label{fig:Fig2bis}
\end{figure*}

To test for asphericity in the Mn spin density, the MEM reconstruction was
performed using a deliberately unfavorable, non-uniform prior~\cite{papoular}. This
prior consists of spherically symmetric magnetization densities centered on the
Mn and F sites, constructed analytically from the refined moments and the radial
integrals obtained in the least-squares refinement. Because this prior is
strongly biased toward spherical densities, any asphericity emerging in the MEM
solution must be supported by the experimental data.

Figure~\ref{fig:Fig2bis}a,b show the reconstructed three-dimensional
 spin density for the 1~T dataset at 2~K. Despite the
spherical prior, the reconstruction robustly reveals two features: a finite
fluorine spin density and a slight asphericity of the Mn magnetization.

MEM analysis of the 6~T dataset yields a nearly identical spin-density
distribution, apart from the expected overall sign inversion associated with the
domain reversal (see \cite{supp}). Thus, within experimental precision, the application of a 6~T magnetic field does not produce detectable changes in the
magnetization density.

The aspherical component of the Mn moment, $\mu^{a}_{\mathrm{Mn}}$, and the ligand spin
density arising from $p$-$d$ hybridization are expected to be small. By subtracting the
spherical part of the Mn spin density from the MEM reconstruction, we obtain the
deformation density shown in Fig.~\ref{fig:Fig2bis}c. The deformation map confirms the
unbalanced, opposite spin polarization on the ligand sites and reveals, rather than a
pure octupolar pattern, a more complex multipolar distribution with zero net magnetic
moment at the Mn site, yet still compatible with ferroic multipolar ordering.

\subsection{Multipolar Ordering}\label{subsec4}

 The magnetic octupoles that constitute the next term in the magnetic multipole expansion
beyond the dipolar and magnetoelectric multipoles may be induced by an
external magnetic field or arise spontaneously from the internal molecular field
in magnetically ordered compounds. In d-wave altermagnets such as MnF$_2$,
they represent the lowest-order ferroically ordered magnetic quantity and serve
as the natural order parameter for the transition into the altermagnetic
state~\cite{bhowal2024,mcclarty2024}. Density functional calculations
show that the first allowed magnetic multipole in MnF$_2$ with the altermagnetic symmetries is
the ferroic magnetic octupole of $O_{32}$ symmetry, with real-space form
$xy\,m_z$~\cite{bhowal2024}. 

The presence of such an octupole implies a slight deviation of the Mn
magnetization density from spherical symmetry. Because these local deformations
respect the magnetic site symmetry, they cancel when averaged over the Mn
positions.  To quantify the
asphericity, we employ a multipole-expansion formalism.
 We expand the density of an atom  $\rho_{a}$ as a product of a radial
function $R_a(r)$ and real spherical harmonics $Y_{lm}$. Then the  magnetisation density at a given point
is calculated as a sum of $m_a\rho_{a}$ over all magnetic atoms in the unit cell
\begin{equation}
\rho_{a}(\mathbf{r}) =
\sum_{l=0}^{4}
R^{2}_{l,a}(r)
\sum_{m=-l}^{l}
P_{lm,a}\,
Y_{lm}\!\left(\frac{\mathbf{r}}{r}\right),
\end{equation}
where $Y_{lm}$ are normalized real spherical harmonics. 
The local coordinate system is chosen with $z \parallel c$ and $x \parallel [100]$.
Slater-type radial functions were used for Mn and F. The resulting 
magnetic form factor,
\begin{equation}
f_a(\mathbf{k}) =
\int \rho_a(\mathbf{r})\,
e^{i\, \mathbf{k}\cdot\mathbf{r}}\,
d\mathbf{r},
\end{equation}
is then inserted into the general expression for the neutron cross section given in Eq.~\ref{eq:mperp} in the supplementary section.

To identify the symmetry-allowed magnetic multipoles, we analyzed the Mn site symmetry, $(D_{2h})$, generated by the operations $C_{2,[001]}, C_{2,[110]}$, and inversion. Up to (l=4), the invariant basis functions are $z^2$, xy, $35z^4-30z^2+3$, and $2xy(7z^2-1)$, corresponding to the spherical harmonics $Y_{20}$, $Y_{2,-2}$, $Y_{40}$, and $Y_{4,-2}$. The $Y_{2,-2}$ and $Y_{4,-2}$ contributions lead, respectively, to the symmetry-allowed magnetic multipole order parameters $O_{32}\propto xym_z$ and $O_{52}\propto xyz^2m_z$ both of which are constrained to be ferroic by the magneto-crystalline symmetries. In the preceding discussion we have focussed on multipoles with altermagnetic symmetries in $m_z$ but multipoles in other components $m_x, m_y$ are also expected to arise \cite{Buiarelli2025}. The experimental results presented here are not sensitive to these noncollinear multipoles that would also be interesting to probe using polarized neutron scattering.

All collinear symmetry-allowed multipoles were refined against the 1 T and 6 T polarized-neutron diffraction data. The results for the 1 T data set are summarized in Table~\ref{table:mult_models1T}. A refinement including only the spherical Mn magnetic form factor (Model~1) accurately reproduces the \textit{odd} reflections, confirming that these reflections are insensitive to the aspherical component of the magnetization density. In contrast (see below), the \textit{even} reflections require progressively higher-order multipolar terms, demonstrating that they directly probe the anisotropic spin density associated with ferroic multipolar order.

\begin{table}[htbp]
\centering
\caption{Multipolar refinement models for the 2~K, 1~T data set using Slater-type radial functions. 
Agreement factors are given separately for the full odd dataset ($N_\text{peak}=140$) and $\it{observed}$  even ($N_\text{peak}=57$) reflections.}

\begin{tabular}{r S S S S}
\hline
 Model & 1 & 2 & 3 & 4 \\
\hline
$m_\text{Mn} P_{00}$ ($\mu_B$) & -5.17(5) & -5.17(5) & -5.17(5) & -5.18(5) \\
$m_\text{F} P_{00}$ ($\mu_B$)   & 0.0      & 0.0278(10) & 0.0318(10) & 0.0298(12) \\
$m_\text{F} P_{20}$ ($\mu_B$)   & 0.0      & 0.0        & 0.0063(8)    & 0.0033(5) \\
$m_\text{Mn} P_{2-2}$ ($\mu_B$) & 0.0      & 0.0        & 0.0135(8)   & 0.0 \\
$m_\text{Mn} P_{4-2}$ ($\mu_B$) & 0.0      & 0.0        & 0.0         &0.0648(23) \\
\hline
$\chi^2/N_\text{peak}$ (odd)  & 5.85 & 5.80 & 5.78 & 5.79 \\
$\chi^2/N_\text{peak}$ (even) & 101.97 & 9.03 & 3.42 & 1.64 \\
\hline
\end{tabular}
\label{table:mult_models1T}
\end{table}

Consistently with the  initial fits, refinement of the even reflections requires inclusion of the fluorine monopole term, $P_{00}$ (Model~2), demonstrating that a finite magnetic moment on the fluorine atoms is essential for reproducing the observed intensities.

Further improvement in the description of the even reflections is achieved by introducing aspherical magnetization densities. For the magnetic point group, the lowest-order symmetry-allowed basis functions up to $l=2$ are $z^2$ and $xy$, corresponding to the spherical harmonics $Y_{20}$ and $Y_{2-2}$, respectively. Including the theoretically predicted Mn $P_{2-2}$ coefficient (Model~3) reduces the discrepancy for the even reflections by nearly a factor of three, while the additional fluorine $P_{20}$ term provides only a modest further improvement.

Motivated by the MEM reconstruction, which exhibits a magnetization-density deformation resembling a hexadecapolar pattern, we also tested the fourth-order harmonics $Y_{40}$, $Y_{4-2}$, and $Y_{4-4}$. Among these, inclusion of the Mn $P_{4-2}$ coefficient (Model~4), associated with the rank-5 magnetic multipole $O_{52}$, yields the best agreement with experiment, reducing the misfit for the even reflections to $\chi^2/N_{\mathrm{peak}} = 1.64$.

To qualitatively assess the origin of the improved agreement, we estimated the observed and calculated Fourier components of the magnetization density using the approximate relation
\begin{equation}
F_{M,\mathrm{obs(calc)}} \approx \frac{F_N A_{s,\mathrm{obs(calc)}}}{2},
\end{equation}
which applies to even reflections with flipping ratios close to unity. For the odd reflections, for which $R \gg 1$, we instead used
\begin{equation}
F_{M,\mathrm{obs(calc)}} \approx F_N A_{s,\mathrm{obs(calc)}}. 
\end{equation}

\begin{figure}[ht]
\centering
\includegraphics[width=0.8\linewidth]{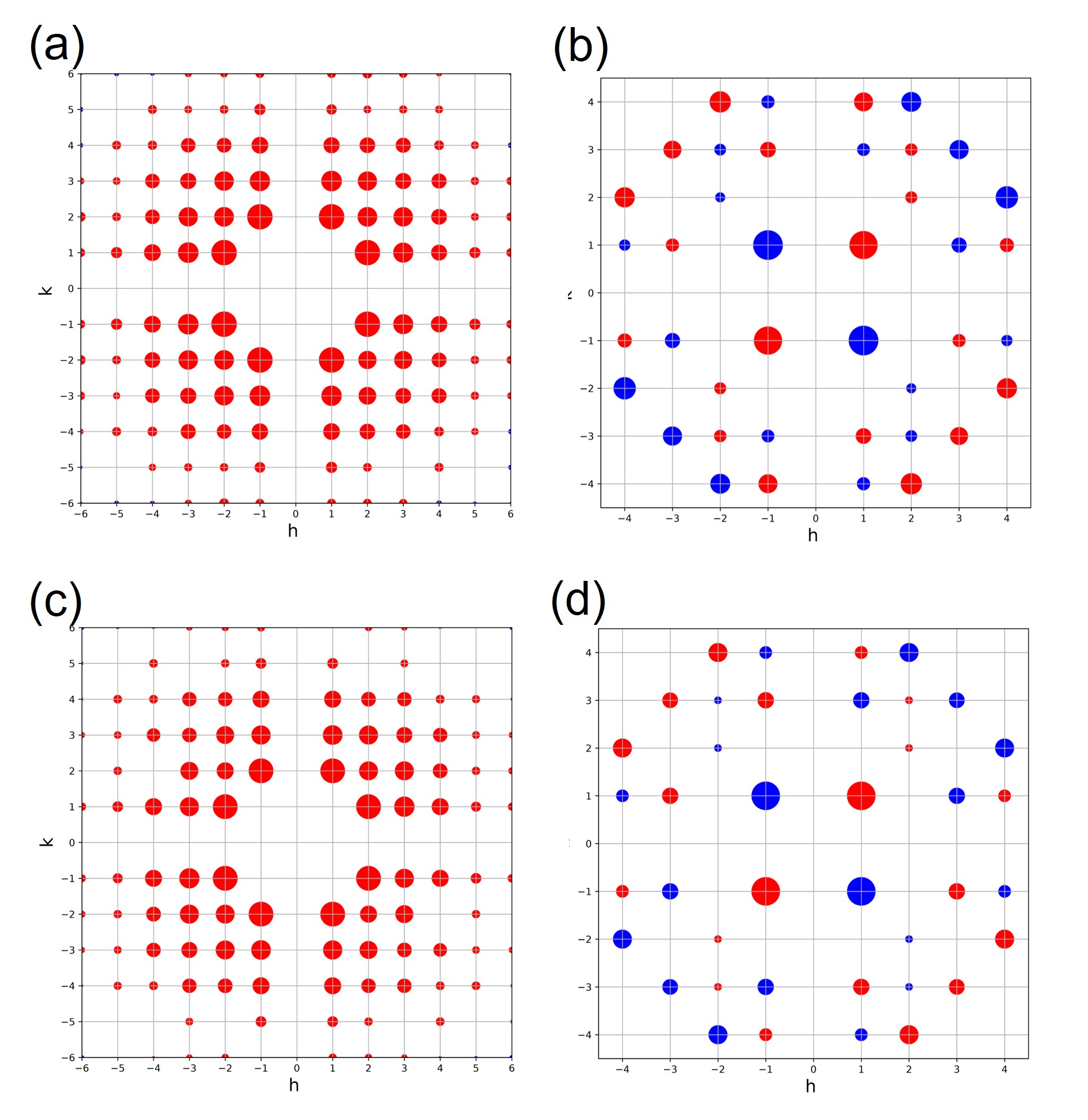}
\caption{Fourier components of the magnetization density. (a,b) Experimental Fourier components derived from the 1 T polarized-neutron diffraction data for the \textit{odd} (a) and \textit{even} (b) reflections. (c,d) Corresponding Fourier components calculated using Model 4 for the \textit{odd} (c) and \textit{even} (d) reflections. The area of each circle is proportional to the magnitude of the corresponding Fourier component. Blue and red circles denote positive and negative Fourier components, respectively. Circle size scale factor 40 is applied to the \textit{even} Fourier  components  for  visibility.
}
\label{fig:Fig3}
\end{figure}

The observed and calculated magnetic Fourier components for the \textit{odd} reflections are compared in Fig.~\ref{fig:Fig3}a,c, where the area of each circle is proportional to the magnitude of the corresponding Fourier component. The excellent agreement between experiment and Model~4 confirms that the spherical component of the Mn magnetization density is accurately reproduced. Consistent with the symmetry of the antiferromagnetic structure, the \textit{odd} Fourier components preserve fourfold rotational symmetry.

By contrast, the Fourier components of the \textit{even} reflections (Fig.~\ref{fig:Fig3}b,d), which probe the aspherical component of the magnetization density, change sign under a $(90^\circ)$ rotation. This alternating sign pattern is the reciprocal-space fingerprint of altermagnetism. Model~4 quantitatively reproduces both the amplitudes and the sign reversal observed experimentally, providing direct evidence for the anisotropic spin density associated with the ferroic multipolar order.

\begin{figure}[t]
\centering
\includegraphics[width=0.8\linewidth]{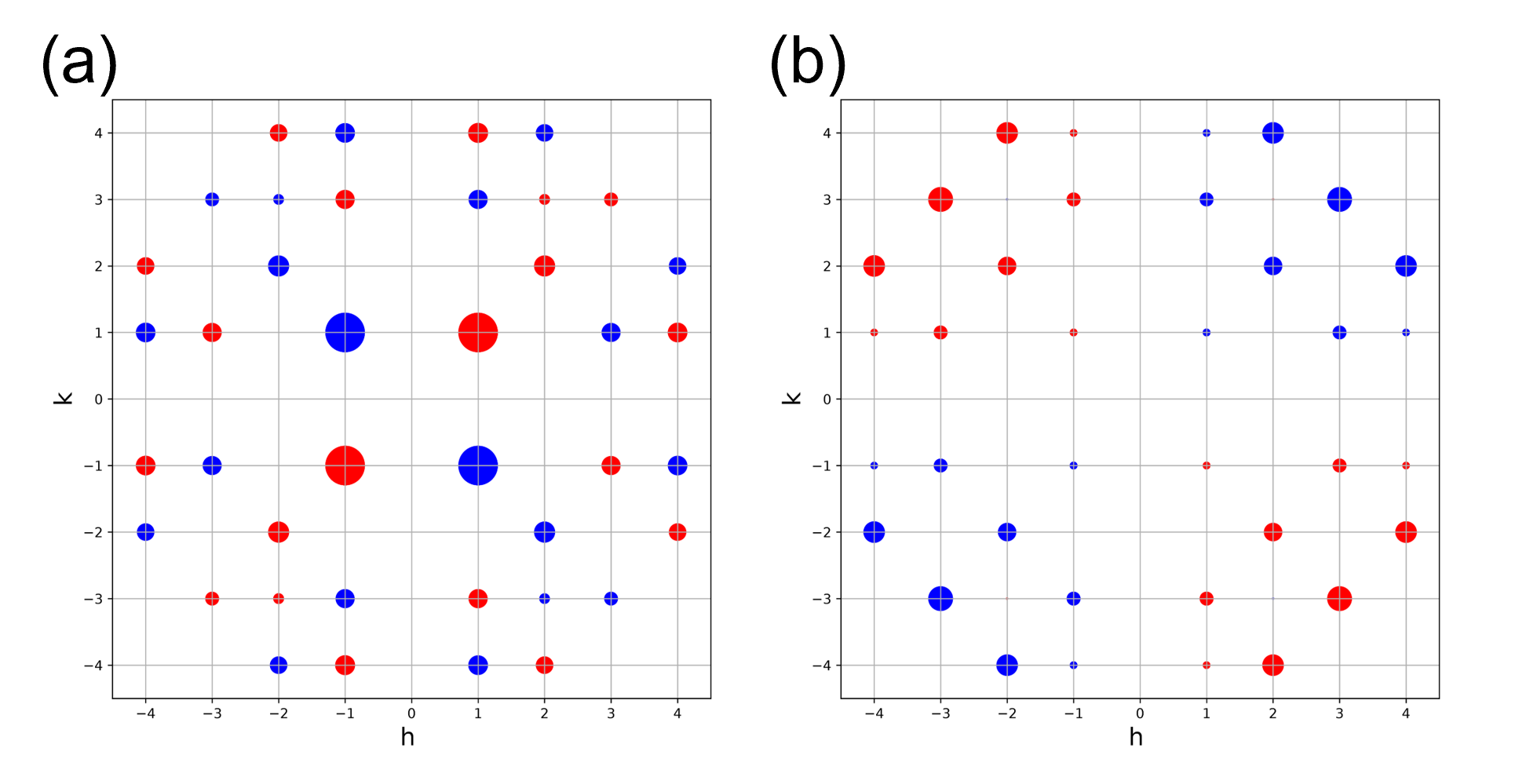}
\caption{
 Fourier components of   calculated magnetization density (Model 4) of the even hkl from 1~T dataset projected onto the (001) plane. (a) Fluorine $m_\text{F} (P_{00} + P_{20})$ and (b) manganese  $m_\text{Mn} P_{4-2} $ components of the density respectively.  
}
\label{fig:Fig4}
\end{figure}

The respective roles of the ligand and multipolar contributions are illustrated in Fig.~\ref{fig:Fig4}, where the Fourier components arising from the fluorine magnetic moment and the Mn $P_{4-2}$ multipole are shown separately. The fluorine contribution accounts remarkably well for the low-index reflections but rapidly decreases with increasing scattering vector and therefore cannot reproduce the high-index reflections. This behavior reflects the rapid decay of the ligand $p$-orbital radial function, further enhanced by Mn--F hybridization. In contrast, the Mn $P_{4-2}$ multipole contributes only weakly at low $q$ but becomes the dominant anisotropic contribution at large scattering vectors, where it is essential for reproducing the observed intensities.

We also tested the other symmetry-allowed $m=0$ multipoles ($P_{20}$ and $P_{40}$), but their inclusion did not improve the refinement, effectively ruling out significant contributions from $m=0$ magnetic multipoles.

The resulting spin-density distribution (Model~4) is shown in Fig.~\ref{fig:Fig2bis}d,e.
Subtracting from the refined model of the spin density the dominant spherical Mn term of $-5.17(4)\,\mu_B$ isolates the deformation spin density shown in Fig.~\ref{fig:Fig2bis}f. The maps reveal ferroic order of the $O_{52}$ rank-5 magnetic multipole (or magnetic triacontadipole) and an imbalance of spin density on the fluorine sites, reflecting time-reversal symmetry breaking of the four-fold screw axis. A close correspondence between this model density and the deformation map obtained from the MEM reconstruction is clearly visible. The fluorine density exhibits a $p_{z}$-like character, although this feature should be interpreted cautiously given the small magnitude of $P_{20}^{\mathrm{F}}$.

\begin{figure}[ht]
\centering
\includegraphics[width=0.8\linewidth]{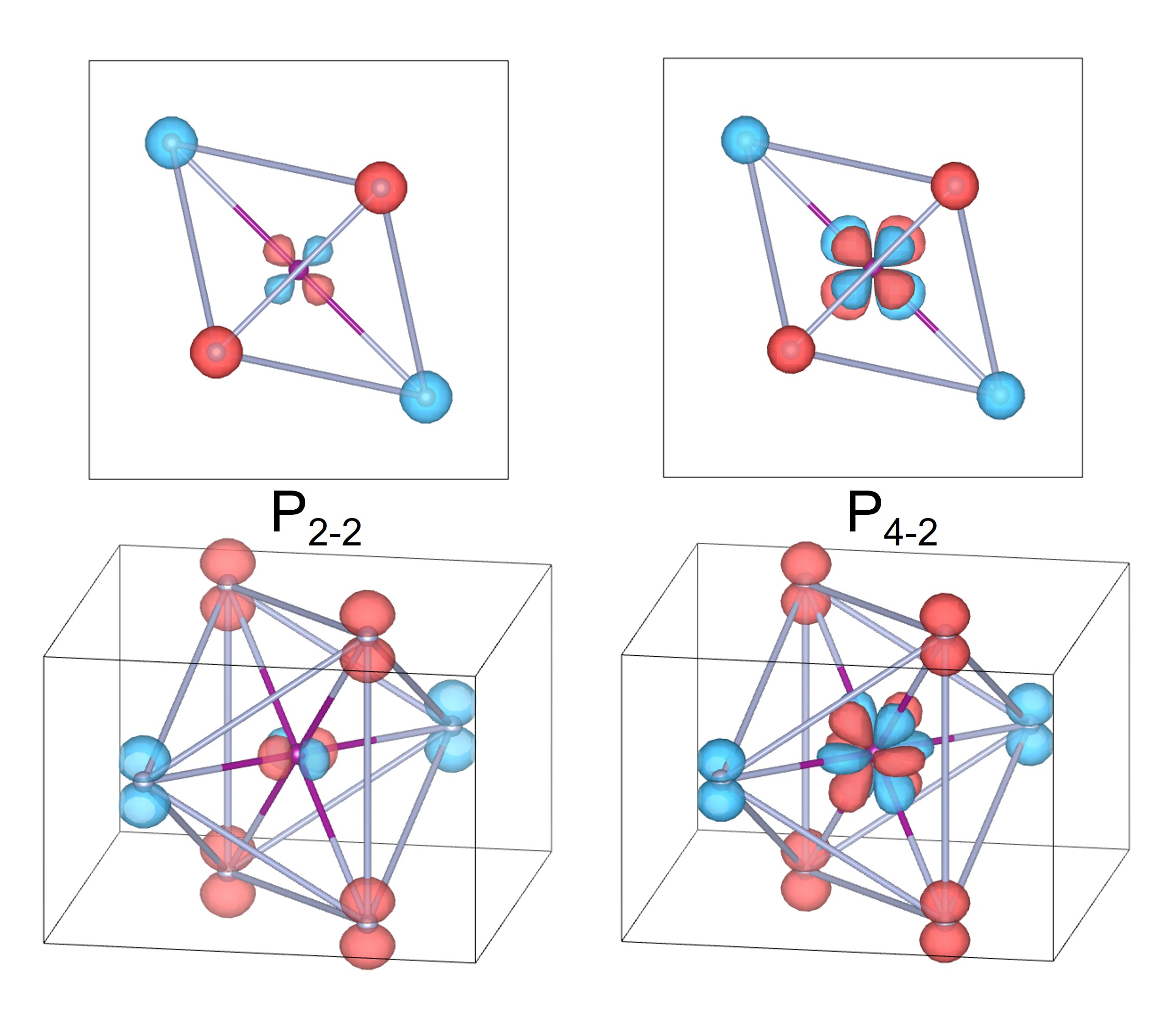}
\caption{
Spin density distribution within the  fluorine cluster combined with  $P_{2-2,\text{Mn}}$ and $ P_{4-2,\text{Mn}}$ multipoles from models 3 and 4, respectively after subtracting the dominant spherical Mn contribution. Top and bottom figures represent projection onto the ab plane and three dimensional spin density distribution, respectively.  The isosurface level of
$0.02\,\mu_B/\text{\AA}^3$ is shown.
}
\label{fig:Fig5}
\end{figure}

While the contribution of the $P_{2-2}$ multipole cannot be totally excluded, beyond the improvement in the fit and the indications from the  MEM reconstruction,  the advantage of the $P_{4-2}$ model is also reflected in the
orientation of its density lobes. In the $P_{4-2}$ case, the lobes point toward
the strongest fluorine bonds (Fig.~\ref{fig:Fig5}), whereas in the $P_{2-2}$ model they lie in the basal plane,
where the basal plane exchange pathway including two fluorine ions is comparatively weak. Experimentally distinguishing between the $P_{2-2}$ and $P_{4-2}$ Mn multipoles would require measurements of reflections with indices $l>1$, which were unfortunately not accessible on the D3 diffractometer.

A similar refinement  performed for the 6~T dataset, has shown that within experimental uncertainty, the multipolar
decomposition remains essentially unchanged (SI).

Thus, both
refinements provide clear evidence for multipoles on the manganese sites incorporating the altermagnetic symmetries. The small
but robust fluorine spin density, antiparallel to the nearest neighbor (nn) Mn moment, is likewise
confirmed. The absence of significant field-induced changes up to 6~T indicates
that the ferro-octupolar order is governed primarily by the Mn molecular field.

\section{Conclusion}\label{sec3}

Polarized-neutron diffraction provides direct real-space evidence of altermagnetic order in MnF$_2$. The measurements reveal a robust covalent spin polarization on the fluorine ligands, $(\mu{_\mathrm{F}} = 0.029(1),\mu_B)$, antiparallel to the nearest-neighbor Mn moments, and a Mn magnetization density that cannot be described by a purely spherical magnetic form factor. Maximum-entropy reconstructions and symmetry-constrained multipole refinements consistently identify the leading magnetic multipoles permitted by the $(D_{2h})$ site symmetry, including the triacontadipolar component $(O_{52}^{-})$. The absence of measurable changes in the spin density up to 6 T indicates that the ferroic multipolar order is dominated by the internal exchange field rather than by the applied magnetic field. More broadly, our work demonstrates that polarized-neutron diffraction can directly resolve the microscopic real-space signatures of altermagnetism, providing a powerful approach for identifying and quantifying ferroic multipoles in altermagnetic materials.

{\bf Acknowledgements} $-$ We acknowledge financial support from the F\'ed\'eration
Fran\c{c}aise de Diffusion Neutronique (2FDN). PM acknowledges funding from the CNRS and Springboard grant (ANR-21-EXE5-0003).

\bibliography{MnF2}

\begin{thebibliography}{39}%
\makeatletter
\providecommand \@ifxundefined [1]{%
 \@ifx{#1\undefined}
}%
\providecommand \@ifnum [1]{%
 \ifnum #1\expandafter \@firstoftwo
 \else \expandafter \@secondoftwo
 \fi
}%
\providecommand \@ifx [1]{%
 \ifx #1\expandafter \@firstoftwo
 \else \expandafter \@secondoftwo
 \fi
}%
\providecommand \natexlab [1]{#1}%
\providecommand \enquote  [1]{``#1''}%
\providecommand \bibnamefont  [1]{#1}%
\providecommand \bibfnamefont [1]{#1}%
\providecommand \citenamefont [1]{#1}%
\providecommand \href@noop [0]{\@secondoftwo}%
\providecommand \href [0]{\begingroup \@sanitize@url \@href}%
\providecommand \@href[1]{\@@startlink{#1}\@@href}%
\providecommand \@@href[1]{\endgroup#1\@@endlink}%
\providecommand \@sanitize@url [0]{\catcode `\\12\catcode `\$12\catcode
  `\&12\catcode `\#12\catcode `\^12\catcode `\_12\catcode `\%12\relax}%
\providecommand \@@startlink[1]{}%
\providecommand \@@endlink[0]{}%
\providecommand \url  [0]{\begingroup\@sanitize@url \@url }%
\providecommand \@url [1]{\endgroup\@href {#1}{\urlprefix }}%
\providecommand \urlprefix  [0]{URL }%
\providecommand \Eprint [0]{\href }%
\providecommand \doibase [0]{https://doi.org/}%
\providecommand \selectlanguage [0]{\@gobble}%
\providecommand \bibinfo  [0]{\@secondoftwo}%
\providecommand \bibfield  [0]{\@secondoftwo}%
\providecommand \translation [1]{[#1]}%
\providecommand \BibitemOpen [0]{}%
\providecommand \bibitemStop [0]{}%
\providecommand \bibitemNoStop [0]{.\EOS\space}%
\providecommand \EOS [0]{\spacefactor3000\relax}%
\providecommand \BibitemShut  [1]{\csname bibitem#1\endcsname}%
\let\auto@bib@innerbib\@empty
\bibitem [{\citenamefont {{Naka}}\ \emph {et~al.}(2019)\citenamefont {{Naka}},
  \citenamefont {{Hayami}}, \citenamefont {{Kusunose}}, \citenamefont
  {{Yanagi}}, \citenamefont {{Motome}},\ and\ \citenamefont
  {{Seo}}}]{naka2019}%
  \BibitemOpen
  \bibfield  {author} {\bibinfo {author} {\bibfnamefont {M.}~\bibnamefont
  {{Naka}}}, \bibinfo {author} {\bibfnamefont {S.}~\bibnamefont {{Hayami}}},
  \bibinfo {author} {\bibfnamefont {H.}~\bibnamefont {{Kusunose}}}, \bibinfo
  {author} {\bibfnamefont {Y.}~\bibnamefont {{Yanagi}}}, \bibinfo {author}
  {\bibfnamefont {Y.}~\bibnamefont {{Motome}}},\ and\ \bibinfo {author}
  {\bibfnamefont {H.}~\bibnamefont {{Seo}}},\ }\bibfield  {title} {\bibinfo
  {title} {{Spin current generation in organic antiferromagnets}},\ }\href
  {https://doi.org/10.1038/s41467-019-12229-y} {\bibfield  {journal} {\bibinfo
  {journal} {Nature Communications}\ }\textbf {\bibinfo {volume} {10}},\
  \bibinfo {eid} {4305} (\bibinfo {year} {2019})},\ \Eprint
  {https://arxiv.org/abs/1902.02506} {arXiv:1902.02506 [cond-mat.str-el]}
  \BibitemShut {NoStop}%
\bibitem [{\citenamefont {{{\v{S}}mejkal}}\ \emph {et~al.}(2020)\citenamefont
  {{{\v{S}}mejkal}}, \citenamefont {{Gonz{\'a}lez-Hern{\'a}ndez}},
  \citenamefont {{Jungwirth}},\ and\ \citenamefont {{Sinova}}}]{Smejkal2019}%
  \BibitemOpen
  \bibfield  {author} {\bibinfo {author} {\bibfnamefont {L.}~\bibnamefont
  {{{\v{S}}mejkal}}}, \bibinfo {author} {\bibfnamefont {R.}~\bibnamefont
  {{Gonz{\'a}lez-Hern{\'a}ndez}}}, \bibinfo {author} {\bibfnamefont
  {T.}~\bibnamefont {{Jungwirth}}},\ and\ \bibinfo {author} {\bibfnamefont
  {J.}~\bibnamefont {{Sinova}}},\ }\bibfield  {title} {\bibinfo {title}
  {{Crystal time-reversal symmetry breaking and spontaneous Hall effect in
  collinear antiferromagnets}},\ }\href
  {https://doi.org/10.1126/sciadv.aaz8809} {\bibfield  {journal} {\bibinfo
  {journal} {Science Advances}\ }\textbf {\bibinfo {volume} {6}},\ \bibinfo
  {pages} {eaaz8809} (\bibinfo {year} {2020})}\BibitemShut {NoStop}%
\bibitem [{\citenamefont {{Hayami}}\ \emph {et~al.}(2019)\citenamefont
  {{Hayami}}, \citenamefont {{Yanagi}},\ and\ \citenamefont
  {{Kusunose}}}]{hayami2019}%
  \BibitemOpen
  \bibfield  {author} {\bibinfo {author} {\bibfnamefont {S.}~\bibnamefont
  {{Hayami}}}, \bibinfo {author} {\bibfnamefont {Y.}~\bibnamefont {{Yanagi}}},\
  and\ \bibinfo {author} {\bibfnamefont {H.}~\bibnamefont {{Kusunose}}},\
  }\bibfield  {title} {\bibinfo {title} {{Momentum-Dependent Spin Splitting by
  Collinear Antiferromagnetic Ordering}},\ }\href
  {https://doi.org/10.7566/JPSJ.88.123702} {\bibfield  {journal} {\bibinfo
  {journal} {Journal of the Physical Society of Japan}\ }\textbf {\bibinfo
  {volume} {88}},\ \bibinfo {eid} {123702} (\bibinfo {year} {2019})},\ \Eprint
  {https://arxiv.org/abs/1908.08680} {arXiv:1908.08680 [cond-mat.str-el]}
  \BibitemShut {NoStop}%
\bibitem [{\citenamefont {Yuan}\ \emph {et~al.}(2020)\citenamefont {Yuan},
  \citenamefont {Wang}, \citenamefont {Luo}, \citenamefont {Rashba},\ and\
  \citenamefont {Zunger}}]{yuan2020}%
  \BibitemOpen
  \bibfield  {author} {\bibinfo {author} {\bibfnamefont {L.-D.}\ \bibnamefont
  {Yuan}}, \bibinfo {author} {\bibfnamefont {Z.}~\bibnamefont {Wang}}, \bibinfo
  {author} {\bibfnamefont {J.-W.}\ \bibnamefont {Luo}}, \bibinfo {author}
  {\bibfnamefont {E.~I.}\ \bibnamefont {Rashba}},\ and\ \bibinfo {author}
  {\bibfnamefont {A.}~\bibnamefont {Zunger}},\ }\bibfield  {title} {\bibinfo
  {title} {Giant momentum-dependent spin splitting in centrosymmetric low-$z$
  antiferromagnets},\ }\href {https://doi.org/10.1103/PhysRevB.102.014422}
  {\bibfield  {journal} {\bibinfo  {journal} {Phys. Rev. B}\ }\textbf {\bibinfo
  {volume} {102}},\ \bibinfo {pages} {014422} (\bibinfo {year}
  {2020})}\BibitemShut {NoStop}%
\bibitem [{\citenamefont {\v{S}mejkal}\ \emph {et~al.}(2023)\citenamefont
  {\v{S}mejkal}, \citenamefont {Marmodoro}, \citenamefont {Ahn}, \citenamefont
  {Gonz\'alez-Hern\'andez}, \citenamefont {Turek}, \citenamefont {Mankovsky},
  \citenamefont {Ebert}, \citenamefont {D'Souza}, \citenamefont {\v{S}ipr},
  \citenamefont {Sinova},\ and\ \citenamefont {Jungwirth}}]{Smejkal2023}%
  \BibitemOpen
  \bibfield  {author} {\bibinfo {author} {\bibfnamefont {L.}~\bibnamefont
  {\v{S}mejkal}}, \bibinfo {author} {\bibfnamefont {A.}~\bibnamefont
  {Marmodoro}}, \bibinfo {author} {\bibfnamefont {K.-H.}\ \bibnamefont {Ahn}},
  \bibinfo {author} {\bibfnamefont {R.}~\bibnamefont {Gonz\'alez-Hern\'andez}},
  \bibinfo {author} {\bibfnamefont {I.}~\bibnamefont {Turek}}, \bibinfo
  {author} {\bibfnamefont {S.}~\bibnamefont {Mankovsky}}, \bibinfo {author}
  {\bibfnamefont {H.}~\bibnamefont {Ebert}}, \bibinfo {author} {\bibfnamefont
  {S.~W.}\ \bibnamefont {D'Souza}}, \bibinfo {author} {\bibfnamefont
  {O.}~\bibnamefont {\v{S}ipr}}, \bibinfo {author} {\bibfnamefont
  {J.}~\bibnamefont {Sinova}},\ and\ \bibinfo {author} {\bibfnamefont
  {T.}~\bibnamefont {Jungwirth}},\ }\bibfield  {title} {\bibinfo {title}
  {Chiral magnons in altermagnetic {${\mathrm{RuO}}_{2}$}},\ }\href
  {https://doi.org/10.1103/PhysRevLett.131.256703} {\bibfield  {journal}
  {\bibinfo  {journal} {Phys. Rev. Lett.}\ }\textbf {\bibinfo {volume} {131}},\
  \bibinfo {pages} {256703} (\bibinfo {year} {2023})}\BibitemShut {NoStop}%
\bibitem [{\citenamefont {\v{S}mejkal}\ \emph {et~al.}(2022)\citenamefont
  {\v{S}mejkal}, \citenamefont {Sinova},\ and\ \citenamefont
  {Jungwirth}}]{Smejkal2022b}%
  \BibitemOpen
  \bibfield  {author} {\bibinfo {author} {\bibfnamefont {L.}~\bibnamefont
  {\v{S}mejkal}}, \bibinfo {author} {\bibfnamefont {J.}~\bibnamefont
  {Sinova}},\ and\ \bibinfo {author} {\bibfnamefont {T.}~\bibnamefont
  {Jungwirth}},\ }\bibfield  {title} {\bibinfo {title} {{Emerging Research
  Landscape of Altermagnetism}},\ }\href
  {https://doi.org/10.1103/PhysRevX.12.040501} {\bibfield  {journal} {\bibinfo
  {journal} {Phys. Rev. X}\ }\textbf {\bibinfo {volume} {12}},\ \bibinfo
  {pages} {040501} (\bibinfo {year} {2022})}\BibitemShut {NoStop}%
\bibitem [{\citenamefont {{Jungwirth}}\ \emph {et~al.}(2025)\citenamefont
  {{Jungwirth}}, \citenamefont {{Sinova}}, \citenamefont {{Fernandes}},
  \citenamefont {{Liu}}, \citenamefont {{Watanabe}}, \citenamefont
  {{Murakami}}, \citenamefont {{Nakatsuji}},\ and\ \citenamefont
  {{Smejkal}}}]{jungwirth2025review}%
  \BibitemOpen
  \bibfield  {author} {\bibinfo {author} {\bibfnamefont {T.}~\bibnamefont
  {{Jungwirth}}}, \bibinfo {author} {\bibfnamefont {J.}~\bibnamefont
  {{Sinova}}}, \bibinfo {author} {\bibfnamefont {R.~M.}\ \bibnamefont
  {{Fernandes}}}, \bibinfo {author} {\bibfnamefont {Q.}~\bibnamefont {{Liu}}},
  \bibinfo {author} {\bibfnamefont {H.}~\bibnamefont {{Watanabe}}}, \bibinfo
  {author} {\bibfnamefont {S.}~\bibnamefont {{Murakami}}}, \bibinfo {author}
  {\bibfnamefont {S.}~\bibnamefont {{Nakatsuji}}},\ and\ \bibinfo {author}
  {\bibfnamefont {L.}~\bibnamefont {{Smejkal}}},\ }\bibfield  {title} {\bibinfo
  {title} {{Symmetry, microscopy and spectroscopy signatures of
  altermagnetism}},\ }\href {https://doi.org/10.48550/arXiv.2506.22860}
  {\bibfield  {journal} {\bibinfo  {journal} {arXiv e-prints}\ ,\ \bibinfo
  {eid} {arXiv:2506.22860}} (\bibinfo {year} {2025})},\ \Eprint
  {https://arxiv.org/abs/2506.22860} {arXiv:2506.22860 [cond-mat.mtrl-sci]}
  \BibitemShut {NoStop}%
\bibitem [{\citenamefont {{Gonzalez Betancourt}}\ \emph
  {et~al.}(2023)\citenamefont {{Gonzalez Betancourt}}, \citenamefont
  {{Zub{\'a}{\v{c}}}}, \citenamefont {{Gonzalez-Hernandez}}, \citenamefont
  {{Geishendorf}}, \citenamefont {{{\v{S}}ob{\'a}{\r{A}}}}, \citenamefont
  {{Springholz}}, \citenamefont {{Olejn{\'\i}k}}, \citenamefont
  {{{\v{S}}mejkal}}, \citenamefont {{Sinova}}, \citenamefont {{Jungwirth}},
  \citenamefont {{Goennenwein}}, \citenamefont {{Thomas}}, \citenamefont
  {{Reichlov{\'a}}}, \citenamefont {{{\v{Z}}elezn{\'y}}},\ and\ \citenamefont
  {{Kriegner}}}]{gonzalez}%
  \BibitemOpen
  \bibfield  {author} {\bibinfo {author} {\bibfnamefont {R.~D.}\ \bibnamefont
  {{Gonzalez Betancourt}}}, \bibinfo {author} {\bibfnamefont {J.}~\bibnamefont
  {{Zub{\'a}{\v{c}}}}}, \bibinfo {author} {\bibfnamefont {R.}~\bibnamefont
  {{Gonzalez-Hernandez}}}, \bibinfo {author} {\bibfnamefont {K.}~\bibnamefont
  {{Geishendorf}}}, \bibinfo {author} {\bibfnamefont {Z.}~\bibnamefont
  {{{\v{S}}ob{\'a}{\r{A}}}}}, \bibinfo {author} {\bibfnamefont
  {G.}~\bibnamefont {{Springholz}}}, \bibinfo {author} {\bibfnamefont
  {K.}~\bibnamefont {{Olejn{\'\i}k}}}, \bibinfo {author} {\bibfnamefont
  {L.}~\bibnamefont {{{\v{S}}mejkal}}}, \bibinfo {author} {\bibfnamefont
  {J.}~\bibnamefont {{Sinova}}}, \bibinfo {author} {\bibfnamefont
  {T.}~\bibnamefont {{Jungwirth}}}, \bibinfo {author} {\bibfnamefont
  {S.~T.~B.}\ \bibnamefont {{Goennenwein}}}, \bibinfo {author} {\bibfnamefont
  {A.}~\bibnamefont {{Thomas}}}, \bibinfo {author} {\bibfnamefont
  {H.}~\bibnamefont {{Reichlov{\'a}}}}, \bibinfo {author} {\bibfnamefont
  {J.}~\bibnamefont {{{\v{Z}}elezn{\'y}}}},\ and\ \bibinfo {author}
  {\bibfnamefont {D.}~\bibnamefont {{Kriegner}}},\ }\bibfield  {title}
  {\bibinfo {title} {{Spontaneous Anomalous Hall Effect Arising from an
  Unconventional Compensated Magnetic Phase in a Semiconductor}},\ }\href
  {https://doi.org/10.1103/PhysRevLett.130.036702} {\bibfield  {journal}
  {\bibinfo  {journal} {Phys. Rev. Lett.}\ }\textbf {\bibinfo {volume} {130}},\
  \bibinfo {eid} {036702} (\bibinfo {year} {2023})},\ \Eprint
  {https://arxiv.org/abs/2112.06805} {arXiv:2112.06805 [cond-mat.mtrl-sci]}
  \BibitemShut {NoStop}%
\bibitem [{\citenamefont {Lee}\ \emph {et~al.}(2024)\citenamefont {Lee},
  \citenamefont {Lee}, \citenamefont {Jung}, \citenamefont {Jung},
  \citenamefont {Kim}, \citenamefont {Lee}, \citenamefont {Seok}, \citenamefont
  {Kim}, \citenamefont {Park}, \citenamefont {\v{S}mejkal}, \citenamefont
  {Kang},\ and\ \citenamefont {Kim}}]{lee2024}%
  \BibitemOpen
  \bibfield  {author} {\bibinfo {author} {\bibfnamefont {S.}~\bibnamefont
  {Lee}}, \bibinfo {author} {\bibfnamefont {S.}~\bibnamefont {Lee}}, \bibinfo
  {author} {\bibfnamefont {S.}~\bibnamefont {Jung}}, \bibinfo {author}
  {\bibfnamefont {J.}~\bibnamefont {Jung}}, \bibinfo {author} {\bibfnamefont
  {D.}~\bibnamefont {Kim}}, \bibinfo {author} {\bibfnamefont {Y.}~\bibnamefont
  {Lee}}, \bibinfo {author} {\bibfnamefont {B.}~\bibnamefont {Seok}}, \bibinfo
  {author} {\bibfnamefont {J.}~\bibnamefont {Kim}}, \bibinfo {author}
  {\bibfnamefont {B.~G.}\ \bibnamefont {Park}}, \bibinfo {author}
  {\bibfnamefont {L.}~\bibnamefont {\v{S}mejkal}}, \bibinfo {author}
  {\bibfnamefont {C.-J.}\ \bibnamefont {Kang}},\ and\ \bibinfo {author}
  {\bibfnamefont {C.}~\bibnamefont {Kim}},\ }\bibfield  {title} {\bibinfo
  {title} {{Broken Kramers Degeneracy in Altermagnetic MnTe}},\ }\href
  {https://doi.org/10.1103/PhysRevLett.132.036702} {\bibfield  {journal}
  {\bibinfo  {journal} {Phys. Rev. Lett.}\ }\textbf {\bibinfo {volume} {132}},\
  \bibinfo {pages} {036702} (\bibinfo {year} {2024})}\BibitemShut {NoStop}%
\bibitem [{\citenamefont {Liu}\ \emph {et~al.}(2024)\citenamefont {Liu},
  \citenamefont {Ozeki}, \citenamefont {Asai}, \citenamefont {Itoh},\ and\
  \citenamefont {Masuda}}]{liu2024}%
  \BibitemOpen
  \bibfield  {author} {\bibinfo {author} {\bibfnamefont {Z.}~\bibnamefont
  {Liu}}, \bibinfo {author} {\bibfnamefont {M.}~\bibnamefont {Ozeki}}, \bibinfo
  {author} {\bibfnamefont {S.}~\bibnamefont {Asai}}, \bibinfo {author}
  {\bibfnamefont {S.}~\bibnamefont {Itoh}},\ and\ \bibinfo {author}
  {\bibfnamefont {T.}~\bibnamefont {Masuda}},\ }\bibfield  {title} {\bibinfo
  {title} {{Chiral Split Magnon in Altermagnetic MnTe}},\ }\href
  {https://doi.org/10.1103/PhysRevLett.133.156702} {\bibfield  {journal}
  {\bibinfo  {journal} {Phys. Rev. Lett.}\ }\textbf {\bibinfo {volume} {133}},\
  \bibinfo {pages} {156702} (\bibinfo {year} {2024})}\BibitemShut {NoStop}%
\bibitem [{\citenamefont {Osumi}\ \emph {et~al.}(2024)\citenamefont {Osumi},
  \citenamefont {Souma}, \citenamefont {Aoyama}, \citenamefont {Yamauchi},
  \citenamefont {Honma}, \citenamefont {Nakayama}, \citenamefont {Takahashi},
  \citenamefont {Ohgushi},\ and\ \citenamefont {Sato}}]{osumi2024}%
  \BibitemOpen
  \bibfield  {author} {\bibinfo {author} {\bibfnamefont {T.}~\bibnamefont
  {Osumi}}, \bibinfo {author} {\bibfnamefont {S.}~\bibnamefont {Souma}},
  \bibinfo {author} {\bibfnamefont {T.}~\bibnamefont {Aoyama}}, \bibinfo
  {author} {\bibfnamefont {K.}~\bibnamefont {Yamauchi}}, \bibinfo {author}
  {\bibfnamefont {A.}~\bibnamefont {Honma}}, \bibinfo {author} {\bibfnamefont
  {K.}~\bibnamefont {Nakayama}}, \bibinfo {author} {\bibfnamefont
  {T.}~\bibnamefont {Takahashi}}, \bibinfo {author} {\bibfnamefont
  {K.}~\bibnamefont {Ohgushi}},\ and\ \bibinfo {author} {\bibfnamefont
  {T.}~\bibnamefont {Sato}},\ }\bibfield  {title} {\bibinfo {title}
  {Observation of a giant band splitting in altermagnetic mnte},\ }\href
  {https://doi.org/10.1103/PhysRevB.109.115102} {\bibfield  {journal} {\bibinfo
   {journal} {Phys. Rev. B}\ }\textbf {\bibinfo {volume} {109}},\ \bibinfo
  {pages} {115102} (\bibinfo {year} {2024})}\BibitemShut {NoStop}%
\bibitem [{\citenamefont {{Hajlaoui}}\ \emph {et~al.}(2024)\citenamefont
  {{Hajlaoui}}, \citenamefont {{Wilfred D'Souza}}, \citenamefont
  {{{\v{S}}mejkal}}, \citenamefont {{Kriegner}}, \citenamefont {{Krizman}},
  \citenamefont {{Zakusylo}}, \citenamefont {{Olszowska}}, \citenamefont
  {{Caha}}, \citenamefont {{Michali{\v{c}}ka}}, \citenamefont
  {{S{\'a}nchez-Barriga}}, \citenamefont {{Marmodoro}}, \citenamefont
  {{V{\'y}born{\'y}}}, \citenamefont {{Ernst}}, \citenamefont {{Cinchetti}},
  \citenamefont {{Minar}}, \citenamefont {{Jungwirth}},\ and\ \citenamefont
  {{Springholz}}}]{Hajlaoui}%
  \BibitemOpen
  \bibfield  {author} {\bibinfo {author} {\bibfnamefont {M.}~\bibnamefont
  {{Hajlaoui}}}, \bibinfo {author} {\bibfnamefont {S.}~\bibnamefont {{Wilfred
  D'Souza}}}, \bibinfo {author} {\bibfnamefont {L.}~\bibnamefont
  {{{\v{S}}mejkal}}}, \bibinfo {author} {\bibfnamefont {D.}~\bibnamefont
  {{Kriegner}}}, \bibinfo {author} {\bibfnamefont {G.}~\bibnamefont
  {{Krizman}}}, \bibinfo {author} {\bibfnamefont {T.}~\bibnamefont
  {{Zakusylo}}}, \bibinfo {author} {\bibfnamefont {N.}~\bibnamefont
  {{Olszowska}}}, \bibinfo {author} {\bibfnamefont {O.}~\bibnamefont {{Caha}}},
  \bibinfo {author} {\bibfnamefont {J.}~\bibnamefont {{Michali{\v{c}}ka}}},
  \bibinfo {author} {\bibfnamefont {J.}~\bibnamefont {{S{\'a}nchez-Barriga}}},
  \bibinfo {author} {\bibfnamefont {A.}~\bibnamefont {{Marmodoro}}}, \bibinfo
  {author} {\bibfnamefont {K.}~\bibnamefont {{V{\'y}born{\'y}}}}, \bibinfo
  {author} {\bibfnamefont {A.}~\bibnamefont {{Ernst}}}, \bibinfo {author}
  {\bibfnamefont {M.}~\bibnamefont {{Cinchetti}}}, \bibinfo {author}
  {\bibfnamefont {J.}~\bibnamefont {{Minar}}}, \bibinfo {author} {\bibfnamefont
  {T.}~\bibnamefont {{Jungwirth}}},\ and\ \bibinfo {author} {\bibfnamefont
  {G.}~\bibnamefont {{Springholz}}},\ }\bibfield  {title} {\bibinfo {title}
  {{Temperature Dependence of Relativistic Valence Band Splitting Induced by an
  Altermagnetic Phase Transition}},\ }\href
  {https://doi.org/10.1002/adma.202314076} {\bibfield  {journal} {\bibinfo
  {journal} {Advanced Materials}\ }\textbf {\bibinfo {volume} {36}},\ \bibinfo
  {eid} {2314076} (\bibinfo {year} {2024})},\ \Eprint
  {https://arxiv.org/abs/2401.09187} {arXiv:2401.09187 [cond-mat.mtrl-sci]}
  \BibitemShut {NoStop}%
\bibitem [{\citenamefont {{Reimers}}\ \emph {et~al.}(2024)\citenamefont
  {{Reimers}}, \citenamefont {{Odenbreit}}, \citenamefont {{{\v{S}}mejkal}},
  \citenamefont {{Strocov}}, \citenamefont {{Constantinou}}, \citenamefont
  {{Hellenes}}, \citenamefont {{Jaeschke Ubiergo}}, \citenamefont {{Campos}},
  \citenamefont {{Bharadwaj}}, \citenamefont {{Chakraborty}}, \citenamefont
  {{Denneulin}}, \citenamefont {{Shi}}, \citenamefont {{Dunin-Borkowski}},
  \citenamefont {{Das}}, \citenamefont {{Kl{\"a}ui}}, \citenamefont
  {{Sinova}},\ and\ \citenamefont {{Jourdan}}}]{reimers2024}%
  \BibitemOpen
  \bibfield  {author} {\bibinfo {author} {\bibfnamefont {S.}~\bibnamefont
  {{Reimers}}}, \bibinfo {author} {\bibfnamefont {L.}~\bibnamefont
  {{Odenbreit}}}, \bibinfo {author} {\bibfnamefont {L.}~\bibnamefont
  {{{\v{S}}mejkal}}}, \bibinfo {author} {\bibfnamefont {V.~N.}\ \bibnamefont
  {{Strocov}}}, \bibinfo {author} {\bibfnamefont {P.}~\bibnamefont
  {{Constantinou}}}, \bibinfo {author} {\bibfnamefont {A.~B.}\ \bibnamefont
  {{Hellenes}}}, \bibinfo {author} {\bibfnamefont {R.}~\bibnamefont {{Jaeschke
  Ubiergo}}}, \bibinfo {author} {\bibfnamefont {W.~H.}\ \bibnamefont
  {{Campos}}}, \bibinfo {author} {\bibfnamefont {V.~K.}\ \bibnamefont
  {{Bharadwaj}}}, \bibinfo {author} {\bibfnamefont {A.}~\bibnamefont
  {{Chakraborty}}}, \bibinfo {author} {\bibfnamefont {T.}~\bibnamefont
  {{Denneulin}}}, \bibinfo {author} {\bibfnamefont {W.}~\bibnamefont {{Shi}}},
  \bibinfo {author} {\bibfnamefont {R.~E.}\ \bibnamefont {{Dunin-Borkowski}}},
  \bibinfo {author} {\bibfnamefont {S.}~\bibnamefont {{Das}}}, \bibinfo
  {author} {\bibfnamefont {M.}~\bibnamefont {{Kl{\"a}ui}}}, \bibinfo {author}
  {\bibfnamefont {J.}~\bibnamefont {{Sinova}}},\ and\ \bibinfo {author}
  {\bibfnamefont {M.}~\bibnamefont {{Jourdan}}},\ }\bibfield  {title} {\bibinfo
  {title} {{Direct observation of altermagnetic band splitting in CrSb thin
  films}},\ }\href {https://doi.org/10.1038/s41467-024-46476-5} {\bibfield
  {journal} {\bibinfo  {journal} {Nature Communications}\ }\textbf {\bibinfo
  {volume} {15}},\ \bibinfo {eid} {2116} (\bibinfo {year} {2024})},\ \Eprint
  {https://arxiv.org/abs/2310.17280} {arXiv:2310.17280} \BibitemShut {NoStop}%
\bibitem [{\citenamefont {Zeng}\ \emph {et~al.}(2024)\citenamefont {Zeng},
  \citenamefont {Zhu}, \citenamefont {Zhu}, \citenamefont {Liu}, \citenamefont
  {Ma}, \citenamefont {Hao}, \citenamefont {Liu}, \citenamefont {Qu},
  \citenamefont {Yang}, \citenamefont {Jiang}, \citenamefont {Yamagami},
  \citenamefont {Arita}, \citenamefont {Zhang}, \citenamefont {Shao},
  \citenamefont {Dai}, \citenamefont {Shimada}, \citenamefont {Liu},
  \citenamefont {Ye}, \citenamefont {Huang}, \citenamefont {Liu},\ and\
  \citenamefont {Liu}}]{zeng2024}%
  \BibitemOpen
  \bibfield  {author} {\bibinfo {author} {\bibfnamefont {M.}~\bibnamefont
  {Zeng}}, \bibinfo {author} {\bibfnamefont {M.-Y.}\ \bibnamefont {Zhu}},
  \bibinfo {author} {\bibfnamefont {Y.-P.}\ \bibnamefont {Zhu}}, \bibinfo
  {author} {\bibfnamefont {X.-R.}\ \bibnamefont {Liu}}, \bibinfo {author}
  {\bibfnamefont {X.-M.}\ \bibnamefont {Ma}}, \bibinfo {author} {\bibfnamefont
  {Y.-J.}\ \bibnamefont {Hao}}, \bibinfo {author} {\bibfnamefont
  {P.}~\bibnamefont {Liu}}, \bibinfo {author} {\bibfnamefont {G.}~\bibnamefont
  {Qu}}, \bibinfo {author} {\bibfnamefont {Y.}~\bibnamefont {Yang}}, \bibinfo
  {author} {\bibfnamefont {Z.}~\bibnamefont {Jiang}}, \bibinfo {author}
  {\bibfnamefont {K.}~\bibnamefont {Yamagami}}, \bibinfo {author}
  {\bibfnamefont {M.}~\bibnamefont {Arita}}, \bibinfo {author} {\bibfnamefont
  {X.}~\bibnamefont {Zhang}}, \bibinfo {author} {\bibfnamefont {T.-H.}\
  \bibnamefont {Shao}}, \bibinfo {author} {\bibfnamefont {Y.}~\bibnamefont
  {Dai}}, \bibinfo {author} {\bibfnamefont {K.}~\bibnamefont {Shimada}},
  \bibinfo {author} {\bibfnamefont {Z.}~\bibnamefont {Liu}}, \bibinfo {author}
  {\bibfnamefont {M.}~\bibnamefont {Ye}}, \bibinfo {author} {\bibfnamefont
  {Y.}~\bibnamefont {Huang}}, \bibinfo {author} {\bibfnamefont
  {Q.}~\bibnamefont {Liu}},\ and\ \bibinfo {author} {\bibfnamefont
  {C.}~\bibnamefont {Liu}},\ }\bibfield  {title} {\bibinfo {title} {Observation
  of spin splitting in room-temperature metallic antiferromagnet crsb},\ }\href
  {https://doi.org/https://doi.org/10.1002/advs.202406529} {\bibfield
  {journal} {\bibinfo  {journal} {Advanced Science}\ }\textbf {\bibinfo
  {volume} {11}},\ \bibinfo {pages} {2406529} (\bibinfo {year} {2024})},\
  \Eprint {https://arxiv.org/abs/https://arxiv.org/abs/2405.12679}
  {https://arxiv.org/abs/2405.12679} \BibitemShut {NoStop}%
\bibitem [{\citenamefont {{Yang}}\ \emph {et~al.}(2025)\citenamefont {{Yang}},
  \citenamefont {{Li}}, \citenamefont {{Yang}}, \citenamefont {{Li}},
  \citenamefont {{Zheng}}, \citenamefont {{Zhu}}, \citenamefont {{Pan}},
  \citenamefont {{Xu}}, \citenamefont {{Cao}}, \citenamefont {{Zhao}},
  \citenamefont {{Jana}}, \citenamefont {{Zhang}}, \citenamefont {{Ye}},
  \citenamefont {{Song}}, \citenamefont {{Hu}}, \citenamefont {{Yang}},
  \citenamefont {{Fujii}}, \citenamefont {{Vobornik}}, \citenamefont {{Shi}},
  \citenamefont {{Yuan}}, \citenamefont {{Zhang}}, \citenamefont {{Xu}},\ and\
  \citenamefont {{Liu}}}]{yang2025}%
  \BibitemOpen
  \bibfield  {author} {\bibinfo {author} {\bibfnamefont {G.}~\bibnamefont
  {{Yang}}}, \bibinfo {author} {\bibfnamefont {Z.}~\bibnamefont {{Li}}},
  \bibinfo {author} {\bibfnamefont {S.}~\bibnamefont {{Yang}}}, \bibinfo
  {author} {\bibfnamefont {J.}~\bibnamefont {{Li}}}, \bibinfo {author}
  {\bibfnamefont {H.}~\bibnamefont {{Zheng}}}, \bibinfo {author} {\bibfnamefont
  {W.}~\bibnamefont {{Zhu}}}, \bibinfo {author} {\bibfnamefont
  {Z.}~\bibnamefont {{Pan}}}, \bibinfo {author} {\bibfnamefont
  {Y.}~\bibnamefont {{Xu}}}, \bibinfo {author} {\bibfnamefont {S.}~\bibnamefont
  {{Cao}}}, \bibinfo {author} {\bibfnamefont {W.}~\bibnamefont {{Zhao}}},
  \bibinfo {author} {\bibfnamefont {A.}~\bibnamefont {{Jana}}}, \bibinfo
  {author} {\bibfnamefont {J.}~\bibnamefont {{Zhang}}}, \bibinfo {author}
  {\bibfnamefont {M.}~\bibnamefont {{Ye}}}, \bibinfo {author} {\bibfnamefont
  {Y.}~\bibnamefont {{Song}}}, \bibinfo {author} {\bibfnamefont {L.-H.}\
  \bibnamefont {{Hu}}}, \bibinfo {author} {\bibfnamefont {L.}~\bibnamefont
  {{Yang}}}, \bibinfo {author} {\bibfnamefont {J.}~\bibnamefont {{Fujii}}},
  \bibinfo {author} {\bibfnamefont {I.}~\bibnamefont {{Vobornik}}}, \bibinfo
  {author} {\bibfnamefont {M.}~\bibnamefont {{Shi}}}, \bibinfo {author}
  {\bibfnamefont {H.}~\bibnamefont {{Yuan}}}, \bibinfo {author} {\bibfnamefont
  {Y.}~\bibnamefont {{Zhang}}}, \bibinfo {author} {\bibfnamefont
  {Y.}~\bibnamefont {{Xu}}},\ and\ \bibinfo {author} {\bibfnamefont
  {Y.}~\bibnamefont {{Liu}}},\ }\bibfield  {title} {\bibinfo {title}
  {{Three-dimensional mapping of the altermagnetic spin splitting in CrSb}},\
  }\href {https://doi.org/10.1038/s41467-025-56647-7} {\bibfield  {journal}
  {\bibinfo  {journal} {Nature Communications}\ }\textbf {\bibinfo {volume}
  {16}},\ \bibinfo {eid} {1442} (\bibinfo {year} {2025})},\ \Eprint
  {https://arxiv.org/abs/2405.12575} {arXiv:2405.12575} \BibitemShut {NoStop}%
\bibitem [{\citenamefont {Ding}\ \emph {et~al.}(2024)\citenamefont {Ding},
  \citenamefont {Jiang}, \citenamefont {Chen}, \citenamefont {Tao},
  \citenamefont {Liu}, \citenamefont {Li}, \citenamefont {Liu}, \citenamefont
  {Sun}, \citenamefont {Cheng}, \citenamefont {Liu}, \citenamefont {Yang},
  \citenamefont {Zhang}, \citenamefont {Deng}, \citenamefont {Jing},
  \citenamefont {Huang}, \citenamefont {Shi}, \citenamefont {Ye}, \citenamefont
  {Qiao}, \citenamefont {Wang}, \citenamefont {Guo}, \citenamefont {Feng},\
  and\ \citenamefont {Shen}}]{ding2024}%
  \BibitemOpen
  \bibfield  {author} {\bibinfo {author} {\bibfnamefont {J.}~\bibnamefont
  {Ding}}, \bibinfo {author} {\bibfnamefont {Z.}~\bibnamefont {Jiang}},
  \bibinfo {author} {\bibfnamefont {X.}~\bibnamefont {Chen}}, \bibinfo {author}
  {\bibfnamefont {Z.}~\bibnamefont {Tao}}, \bibinfo {author} {\bibfnamefont
  {Z.}~\bibnamefont {Liu}}, \bibinfo {author} {\bibfnamefont {T.}~\bibnamefont
  {Li}}, \bibinfo {author} {\bibfnamefont {J.}~\bibnamefont {Liu}}, \bibinfo
  {author} {\bibfnamefont {J.}~\bibnamefont {Sun}}, \bibinfo {author}
  {\bibfnamefont {J.}~\bibnamefont {Cheng}}, \bibinfo {author} {\bibfnamefont
  {J.}~\bibnamefont {Liu}}, \bibinfo {author} {\bibfnamefont {Y.}~\bibnamefont
  {Yang}}, \bibinfo {author} {\bibfnamefont {R.}~\bibnamefont {Zhang}},
  \bibinfo {author} {\bibfnamefont {L.}~\bibnamefont {Deng}}, \bibinfo {author}
  {\bibfnamefont {W.}~\bibnamefont {Jing}}, \bibinfo {author} {\bibfnamefont
  {Y.}~\bibnamefont {Huang}}, \bibinfo {author} {\bibfnamefont
  {Y.}~\bibnamefont {Shi}}, \bibinfo {author} {\bibfnamefont {M.}~\bibnamefont
  {Ye}}, \bibinfo {author} {\bibfnamefont {S.}~\bibnamefont {Qiao}}, \bibinfo
  {author} {\bibfnamefont {Y.}~\bibnamefont {Wang}}, \bibinfo {author}
  {\bibfnamefont {Y.}~\bibnamefont {Guo}}, \bibinfo {author} {\bibfnamefont
  {D.}~\bibnamefont {Feng}},\ and\ \bibinfo {author} {\bibfnamefont
  {D.}~\bibnamefont {Shen}},\ }\bibfield  {title} {\bibinfo {title} {Large band
  splitting in $g$-wave altermagnet crsb},\ }\href
  {https://doi.org/10.1103/PhysRevLett.133.206401} {\bibfield  {journal}
  {\bibinfo  {journal} {Phys. Rev. Lett.}\ }\textbf {\bibinfo {volume} {133}},\
  \bibinfo {pages} {206401} (\bibinfo {year} {2024})}\BibitemShut {NoStop}%
\bibitem [{\citenamefont {{Lu}}\ \emph {et~al.}(2025)\citenamefont {{Lu}},
  \citenamefont {{Feng}}, \citenamefont {{Wang}}, \citenamefont {{Chen}},
  \citenamefont {{Lin}}, \citenamefont {{Liang}}, \citenamefont {{Liu}},
  \citenamefont {{Feng}}, \citenamefont {{Yamagami}}, \citenamefont {{Liu}},
  \citenamefont {{Felser}}, \citenamefont {{Wu}},\ and\ \citenamefont
  {{Ma}}}]{lu2024crsb}%
  \BibitemOpen
  \bibfield  {author} {\bibinfo {author} {\bibfnamefont {W.}~\bibnamefont
  {{Lu}}}, \bibinfo {author} {\bibfnamefont {S.}~\bibnamefont {{Feng}}},
  \bibinfo {author} {\bibfnamefont {Y.}~\bibnamefont {{Wang}}}, \bibinfo
  {author} {\bibfnamefont {D.}~\bibnamefont {{Chen}}}, \bibinfo {author}
  {\bibfnamefont {Z.}~\bibnamefont {{Lin}}}, \bibinfo {author} {\bibfnamefont
  {X.}~\bibnamefont {{Liang}}}, \bibinfo {author} {\bibfnamefont
  {S.}~\bibnamefont {{Liu}}}, \bibinfo {author} {\bibfnamefont
  {W.}~\bibnamefont {{Feng}}}, \bibinfo {author} {\bibfnamefont
  {K.}~\bibnamefont {{Yamagami}}}, \bibinfo {author} {\bibfnamefont
  {J.}~\bibnamefont {{Liu}}}, \bibinfo {author} {\bibfnamefont
  {C.}~\bibnamefont {{Felser}}}, \bibinfo {author} {\bibfnamefont
  {Q.}~\bibnamefont {{Wu}}},\ and\ \bibinfo {author} {\bibfnamefont
  {J.}~\bibnamefont {{Ma}}},\ }\bibfield  {title} {\bibinfo {title} {{Signature
  of Topological Surface Bands in Altermagnetic Weyl Semimetal CrSb}},\ }\href
  {https://doi.org/10.1021/acs.nanolett.5c00482} {\bibfield  {journal}
  {\bibinfo  {journal} {Nano Letters}\ }\textbf {\bibinfo {volume} {25}},\
  \bibinfo {pages} {7343} (\bibinfo {year} {2025})},\ \Eprint
  {https://arxiv.org/abs/2407.13497} {arXiv:2407.13497 [cond-mat.mtrl-sci]}
  \BibitemShut {NoStop}%
\bibitem [{\citenamefont {{Faure}}\ \emph {et~al.}(2025)\citenamefont
  {{Faure}}, \citenamefont {{Bounoua}}, \citenamefont {{Bal{\'e}dent}},
  \citenamefont {{Gukasov}}, \citenamefont {{Ovidiu Garlea}}, \citenamefont
  {{Ribeiro}}, \citenamefont {{Rau}}, \citenamefont {{Petit}},\ and\
  \citenamefont {{McClarty}}}]{faure2025}%
  \BibitemOpen
  \bibfield  {author} {\bibinfo {author} {\bibfnamefont {Q.}~\bibnamefont
  {{Faure}}}, \bibinfo {author} {\bibfnamefont {D.}~\bibnamefont {{Bounoua}}},
  \bibinfo {author} {\bibfnamefont {V.}~\bibnamefont {{Bal{\'e}dent}}},
  \bibinfo {author} {\bibfnamefont {A.}~\bibnamefont {{Gukasov}}}, \bibinfo
  {author} {\bibfnamefont {V.}~\bibnamefont {{Ovidiu Garlea}}}, \bibinfo
  {author} {\bibfnamefont {A.}~\bibnamefont {{Ribeiro}}}, \bibinfo {author}
  {\bibfnamefont {J.~G.}\ \bibnamefont {{Rau}}}, \bibinfo {author}
  {\bibfnamefont {S.}~\bibnamefont {{Petit}}},\ and\ \bibinfo {author}
  {\bibfnamefont {P.}~\bibnamefont {{McClarty}}},\ }\bibfield  {title}
  {\bibinfo {title} {{Altermagnetism revealed by polarized neutrons in
  MnF$_2$}},\ }\href {https://doi.org/10.48550/arXiv.2509.07087} {\bibfield
  {journal} {\bibinfo  {journal} {arXiv e-prints}\ ,\ \bibinfo {eid}
  {arXiv:2509.07087}} (\bibinfo {year} {2025})},\ \Eprint
  {https://arxiv.org/abs/2509.07087} {arXiv:2509.07087 [cond-mat.str-el]}
  \BibitemShut {NoStop}%
\bibitem [{\citenamefont {Hayami}\ \emph {et~al.}(2020)\citenamefont {Hayami},
  \citenamefont {Yanagi},\ and\ \citenamefont {Kusunose}}]{hayami2020bottomup}%
  \BibitemOpen
  \bibfield  {author} {\bibinfo {author} {\bibfnamefont {S.}~\bibnamefont
  {Hayami}}, \bibinfo {author} {\bibfnamefont {Y.}~\bibnamefont {Yanagi}},\
  and\ \bibinfo {author} {\bibfnamefont {H.}~\bibnamefont {Kusunose}},\
  }\bibfield  {title} {\bibinfo {title} {Bottom-up design of spin-split and
  reshaped electronic band structures in antiferromagnets without spin-orbit
  coupling: Procedure on the basis of augmented multipoles},\ }\href
  {https://doi.org/10.1103/PhysRevB.102.144441} {\bibfield  {journal} {\bibinfo
   {journal} {Phys. Rev. B}\ }\textbf {\bibinfo {volume} {102}},\ \bibinfo
  {pages} {144441} (\bibinfo {year} {2020})}\BibitemShut {NoStop}%
\bibitem [{\citenamefont {Bhowal}\ and\ \citenamefont
  {Spaldin}(2024)}]{bhowal2024}%
  \BibitemOpen
  \bibfield  {author} {\bibinfo {author} {\bibfnamefont {S.}~\bibnamefont
  {Bhowal}}\ and\ \bibinfo {author} {\bibfnamefont {N.~A.}\ \bibnamefont
  {Spaldin}},\ }\bibfield  {title} {\bibinfo {title} {Ferroically ordered
  magnetic octupoles in $d$-wave altermagnets},\ }\href
  {https://doi.org/10.1103/PhysRevX.14.011019} {\bibfield  {journal} {\bibinfo
  {journal} {Phys. Rev. X}\ }\textbf {\bibinfo {volume} {14}},\ \bibinfo
  {pages} {011019} (\bibinfo {year} {2024})}\BibitemShut {NoStop}%
\bibitem [{\citenamefont {McClarty}\ and\ \citenamefont
  {Rau}(2024)}]{mcclarty2024}%
  \BibitemOpen
  \bibfield  {author} {\bibinfo {author} {\bibfnamefont {P.~A.}\ \bibnamefont
  {McClarty}}\ and\ \bibinfo {author} {\bibfnamefont {J.~G.}\ \bibnamefont
  {Rau}},\ }\bibfield  {title} {\bibinfo {title} {Landau theory of
  altermagnetism},\ }\href {https://doi.org/10.1103/PhysRevLett.132.176702}
  {\bibfield  {journal} {\bibinfo  {journal} {Phys. Rev. Lett.}\ }\textbf
  {\bibinfo {volume} {132}},\ \bibinfo {pages} {176702} (\bibinfo {year}
  {2024})}\BibitemShut {NoStop}%
\bibitem [{\citenamefont {Schiff}\ \emph {et~al.}(2025)\citenamefont {Schiff},
  \citenamefont {McClarty}, \citenamefont {Rau},\ and\ \citenamefont
  {Romh\'anyi}}]{schiff2025}%
  \BibitemOpen
  \bibfield  {author} {\bibinfo {author} {\bibfnamefont {H.}~\bibnamefont
  {Schiff}}, \bibinfo {author} {\bibfnamefont {P.}~\bibnamefont {McClarty}},
  \bibinfo {author} {\bibfnamefont {J.~G.}\ \bibnamefont {Rau}},\ and\ \bibinfo
  {author} {\bibfnamefont {J.}~\bibnamefont {Romh\'anyi}},\ }\bibfield  {title}
  {\bibinfo {title} {Collinear altermagnets and their landau theories},\ }\href
  {https://doi.org/10.1103/q44z-ynbr} {\bibfield  {journal} {\bibinfo
  {journal} {Phys. Rev. Res.}\ }\textbf {\bibinfo {volume} {7}},\ \bibinfo
  {pages} {033301} (\bibinfo {year} {2025})}\BibitemShut {NoStop}%
\bibitem [{\citenamefont {Verbeek}\ \emph {et~al.}(2024)\citenamefont
  {Verbeek}, \citenamefont {Voderholzer}, \citenamefont {Sch\"aren},
  \citenamefont {Gachnang}, \citenamefont {Spaldin},\ and\ \citenamefont
  {Bhowal}}]{verbeek2024}%
  \BibitemOpen
  \bibfield  {author} {\bibinfo {author} {\bibfnamefont {X.~H.}\ \bibnamefont
  {Verbeek}}, \bibinfo {author} {\bibfnamefont {D.}~\bibnamefont
  {Voderholzer}}, \bibinfo {author} {\bibfnamefont {S.}~\bibnamefont
  {Sch\"aren}}, \bibinfo {author} {\bibfnamefont {Y.}~\bibnamefont {Gachnang}},
  \bibinfo {author} {\bibfnamefont {N.~A.}\ \bibnamefont {Spaldin}},\ and\
  \bibinfo {author} {\bibfnamefont {S.}~\bibnamefont {Bhowal}},\ }\bibfield
  {title} {\bibinfo {title} {Nonrelativistic ferromagnetotriakontadipolar order
  and spin splitting in hematite},\ }\href
  {https://doi.org/10.1103/PhysRevResearch.6.043157} {\bibfield  {journal}
  {\bibinfo  {journal} {Phys. Rev. Res.}\ }\textbf {\bibinfo {volume} {6}},\
  \bibinfo {pages} {043157} (\bibinfo {year} {2024})}\BibitemShut {NoStop}%
\bibitem [{\citenamefont {{Costa}}\ and\ \citenamefont
  {{Brown}}(1989)}]{costa1989}%
  \BibitemOpen
  \bibfield  {author} {\bibinfo {author} {\bibfnamefont {M.~M.~R.}\
  \bibnamefont {{Costa}}}\ and\ \bibinfo {author} {\bibfnamefont {P.~J.}\
  \bibnamefont {{Brown}}},\ }\bibfield  {title} {\bibinfo {title}
  {{Magnetisation density in MnF $_{2}$}},\ }\href
  {https://doi.org/10.1016/0921-4526(89)90669-8} {\bibfield  {journal}
  {\bibinfo  {journal} {Physica B Condensed Matter}\ }\textbf {\bibinfo
  {volume} {156}},\ \bibinfo {pages} {329} (\bibinfo {year}
  {1989})}\BibitemShut {NoStop}%
\bibitem [{\citenamefont {Nathans}\ \emph {et~al.}(1963)\citenamefont
  {Nathans}, \citenamefont {Alperin}, \citenamefont {Pickart},\ and\
  \citenamefont {Brown}}]{nathans1963}%
  \BibitemOpen
  \bibfield  {author} {\bibinfo {author} {\bibfnamefont {R.}~\bibnamefont
  {Nathans}}, \bibinfo {author} {\bibfnamefont {H.}~\bibnamefont {Alperin}},
  \bibinfo {author} {\bibfnamefont {S.}~\bibnamefont {Pickart}},\ and\ \bibinfo
  {author} {\bibfnamefont {P.}~\bibnamefont {Brown}},\ }\bibfield  {title}
  {\bibinfo {title} {Measurement of the covalent spin distribution in manganous
  fluoride using polarized neutrons},\ }\href@noop {} {\bibfield  {journal}
  {\bibinfo  {journal} {Journal of Applied Physics}\ }\textbf {\bibinfo
  {volume} {34}},\ \bibinfo {pages} {1182} (\bibinfo {year}
  {1963})}\BibitemShut {NoStop}%
\bibitem [{\citenamefont {Kibalin}\ and\ \citenamefont {Gukasov}()}]{cryspy}%
  \BibitemOpen
  \bibfield  {author} {\bibinfo {author} {\bibfnamefont {I.}~\bibnamefont
  {Kibalin}}\ and\ \bibinfo {author} {\bibfnamefont {A.}~\bibnamefont
  {Gukasov}},\ }\href@noop {} {\bibinfo {title} {Cryspy. crystallographic
  python library}},\ \bibinfo {howpublished}
  {\url{https://www.cryspy.fr/}}\BibitemShut {NoStop}%
\bibitem [{\citenamefont {Jeong}\ \emph {et~al.}(2020)\citenamefont {Jeong},
  \citenamefont {Lenz}, \citenamefont {Gukasov}, \citenamefont {Fabr\`eges},
  \citenamefont {Sazonov}, \citenamefont {Hutanu}, \citenamefont {Louat},
  \citenamefont {Bounoua}, \citenamefont {Martins}, \citenamefont {Biermann},
  \citenamefont {Brouet}, \citenamefont {Sidis},\ and\ \citenamefont
  {Bourges}}]{jeong2020}%
  \BibitemOpen
  \bibfield  {author} {\bibinfo {author} {\bibfnamefont {J.}~\bibnamefont
  {Jeong}}, \bibinfo {author} {\bibfnamefont {B.}~\bibnamefont {Lenz}},
  \bibinfo {author} {\bibfnamefont {A.}~\bibnamefont {Gukasov}}, \bibinfo
  {author} {\bibfnamefont {X.}~\bibnamefont {Fabr\`eges}}, \bibinfo {author}
  {\bibfnamefont {A.}~\bibnamefont {Sazonov}}, \bibinfo {author} {\bibfnamefont
  {V.}~\bibnamefont {Hutanu}}, \bibinfo {author} {\bibfnamefont
  {A.}~\bibnamefont {Louat}}, \bibinfo {author} {\bibfnamefont
  {D.}~\bibnamefont {Bounoua}}, \bibinfo {author} {\bibfnamefont
  {C.}~\bibnamefont {Martins}}, \bibinfo {author} {\bibfnamefont
  {S.}~\bibnamefont {Biermann}}, \bibinfo {author} {\bibfnamefont
  {V.}~\bibnamefont {Brouet}}, \bibinfo {author} {\bibfnamefont
  {Y.}~\bibnamefont {Sidis}},\ and\ \bibinfo {author} {\bibfnamefont
  {P.}~\bibnamefont {Bourges}},\ }\bibfield  {title} {\bibinfo {title}
  {Magnetization density distribution of ${\mathrm{sr}}_{2}{\mathrm{iro}}_{4}$:
  Deviation from a local ${j}_{\mathrm{eff}}=1/2$ picture},\ }\href
  {https://doi.org/10.1103/PhysRevLett.125.097202} {\bibfield  {journal}
  {\bibinfo  {journal} {Phys. Rev. Lett.}\ }\textbf {\bibinfo {volume} {125}},\
  \bibinfo {pages} {097202} (\bibinfo {year} {2020})}\BibitemShut {NoStop}%
\bibitem [{\citenamefont {Alperin}\ \emph {et~al.}(1962)\citenamefont
  {Alperin}, \citenamefont {Brown}, \citenamefont {Nathans},\ and\
  \citenamefont {Pickart}}]{Alperin1962}%
  \BibitemOpen
  \bibfield  {author} {\bibinfo {author} {\bibfnamefont {H.~A.}\ \bibnamefont
  {Alperin}}, \bibinfo {author} {\bibfnamefont {P.~J.}\ \bibnamefont {Brown}},
  \bibinfo {author} {\bibfnamefont {R.}~\bibnamefont {Nathans}},\ and\ \bibinfo
  {author} {\bibfnamefont {S.~J.}\ \bibnamefont {Pickart}},\ }\bibfield
  {title} {\bibinfo {title} {Polarized neutron study of antiferromagnetic
  domains in {Mn${\mathrm{F}}_{2}$}},\ }\href
  {https://doi.org/10.1103/PhysRevLett.8.237} {\bibfield  {journal} {\bibinfo
  {journal} {Phys. Rev. Lett.}\ }\textbf {\bibinfo {volume} {8}},\ \bibinfo
  {pages} {237} (\bibinfo {year} {1962})}\BibitemShut {NoStop}%
\bibitem [{\citenamefont {McClarty}\ \emph {et~al.}(2025)\citenamefont
  {McClarty}, \citenamefont {Gukasov},\ and\ \citenamefont
  {Rau}}]{mcclarty2025}%
  \BibitemOpen
  \bibfield  {author} {\bibinfo {author} {\bibfnamefont {P.~A.}\ \bibnamefont
  {McClarty}}, \bibinfo {author} {\bibfnamefont {A.}~\bibnamefont {Gukasov}},\
  and\ \bibinfo {author} {\bibfnamefont {J.~G.}\ \bibnamefont {Rau}},\
  }\bibfield  {title} {\bibinfo {title} {Observing altermagnetism using
  polarized neutrons},\ }\href {https://doi.org/10.1103/PhysRevB.111.L060405}
  {\bibfield  {journal} {\bibinfo  {journal} {Phys. Rev. B}\ }\textbf {\bibinfo
  {volume} {111}},\ \bibinfo {pages} {L060405} (\bibinfo {year}
  {2025})}\BibitemShut {NoStop}%
\bibitem [{\citenamefont {{Felcher}}\ and\ \citenamefont
  {{Kleb}}(1996)}]{Felcher1996}%
  \BibitemOpen
  \bibfield  {author} {\bibinfo {author} {\bibfnamefont {G.~P.}\ \bibnamefont
  {{Felcher}}}\ and\ \bibinfo {author} {\bibfnamefont {R.}~\bibnamefont
  {{Kleb}}},\ }\bibfield  {title} {\bibinfo {title} {{Antiferromagnetic domains
  and the spin-flop transition of MnF$_{2}$}},\ }\href
  {https://doi.org/10.1209/epl/i1996-00251-7} {\bibfield  {journal} {\bibinfo
  {journal} {EPL (Europhysics Letters)}\ }\textbf {\bibinfo {volume} {36}},\
  \bibinfo {pages} {455} (\bibinfo {year} {1996})}\BibitemShut {NoStop}%
\bibitem [{\citenamefont {Matthewman}()}]{FF}%
  \BibitemOpen
  \bibfield  {author} {\bibinfo {author} {\bibfnamefont {J.~B.~J.}\
  \bibnamefont {Matthewman}},\ }\href@noop {} {\bibinfo {title} {Ccsl,
  ral93-009 (1993).}}\BibitemShut {Stop}%
\bibitem [{\citenamefont {Suzuki}\ \emph {et~al.}(2017)\citenamefont {Suzuki},
  \citenamefont {Koretsune}, \citenamefont {Ochi},\ and\ \citenamefont
  {Arita}}]{suzuki2017}%
  \BibitemOpen
  \bibfield  {author} {\bibinfo {author} {\bibfnamefont {M.-T.}\ \bibnamefont
  {Suzuki}}, \bibinfo {author} {\bibfnamefont {T.}~\bibnamefont {Koretsune}},
  \bibinfo {author} {\bibfnamefont {M.}~\bibnamefont {Ochi}},\ and\ \bibinfo
  {author} {\bibfnamefont {R.}~\bibnamefont {Arita}},\ }\bibfield  {title}
  {\bibinfo {title} {Cluster multipole theory for anomalous hall effect in
  antiferromagnets},\ }\href {https://doi.org/10.1103/PhysRevB.95.094406}
  {\bibfield  {journal} {\bibinfo  {journal} {Phys. Rev. B}\ }\textbf {\bibinfo
  {volume} {95}},\ \bibinfo {pages} {094406} (\bibinfo {year}
  {2017})}\BibitemShut {NoStop}%
\bibitem [{\citenamefont {{Hayami}}\ and\ \citenamefont
  {{Kusunose}}(2024)}]{hayami2024multipole}%
  \BibitemOpen
  \bibfield  {author} {\bibinfo {author} {\bibfnamefont {S.}~\bibnamefont
  {{Hayami}}}\ and\ \bibinfo {author} {\bibfnamefont {H.}~\bibnamefont
  {{Kusunose}}},\ }\bibfield  {title} {\bibinfo {title} {{Unified Description
  of Electronic Orderings and Cross Correlations by Complete Multipole
  Representation}},\ }\href {https://doi.org/10.7566/JPSJ.93.072001} {\bibfield
   {journal} {\bibinfo  {journal} {Journal of the Physical Society of Japan}\
  }\textbf {\bibinfo {volume} {93}},\ \bibinfo {eid} {072001} (\bibinfo {year}
  {2024})},\ \Eprint {https://arxiv.org/abs/2403.09019} {arXiv:2403.09019
  [cond-mat.str-el]} \BibitemShut {NoStop}%
\bibitem [{\citenamefont {{Papoular}}\ and\ \citenamefont
  {{Gillon}}(1990)}]{papoular}%
  \BibitemOpen
  \bibfield  {author} {\bibinfo {author} {\bibfnamefont {R.~J.}\ \bibnamefont
  {{Papoular}}}\ and\ \bibinfo {author} {\bibfnamefont {B.}~\bibnamefont
  {{Gillon}}},\ }\bibfield  {title} {\bibinfo {title} {{Maximum entropy
  reconstruction of spin density maps in crystals from polarized neutron
  diffraction data}},\ }\href {https://doi.org/10.1209/0295-5075/13/5/009}
  {\bibfield  {journal} {\bibinfo  {journal} {EPL (Europhysics Letters)}\
  }\textbf {\bibinfo {volume} {13}},\ \bibinfo {pages} {429} (\bibinfo {year}
  {1990})}\BibitemShut {NoStop}%
\bibitem [{\citenamefont {Karmeshu}(2003)}]{karmeshu2003}%
  \BibitemOpen
  \bibfield  {author} {\bibinfo {author} {\bibfnamefont {J.}~\bibnamefont
  {Karmeshu}},\ }\href@noop {} {\emph {\bibinfo {title} {Entropy Measures,
  Maximum Entropy Principle and Emerging Applications}}}\ (\bibinfo
  {publisher} {Springer-Verlag},\ \bibinfo {year} {2003})\BibitemShut {NoStop}%
\bibitem [{\citenamefont {Kibalin}\ \emph {et~al.}()\citenamefont {Kibalin},
  \citenamefont {Bounoua}, \citenamefont {Velamaz\'an}, \citenamefont {Fabelo},
  \citenamefont {Qureshi}, \citenamefont {Faure}, \citenamefont {Bourges},
  \citenamefont {Bal\'edent}, \citenamefont {Soh}, \citenamefont {Rau},
  \citenamefont {McClarty},\ and\ \citenamefont {Gukasov}}]{supp}%
  \BibitemOpen
  \bibfield  {author} {\bibinfo {author} {\bibfnamefont {I.}~\bibnamefont
  {Kibalin}}, \bibinfo {author} {\bibfnamefont {D.}~\bibnamefont {Bounoua}},
  \bibinfo {author} {\bibfnamefont {J.~A.~R.}\ \bibnamefont {Velamaz\'an}},
  \bibinfo {author} {\bibfnamefont {O.}~\bibnamefont {Fabelo}}, \bibinfo
  {author} {\bibfnamefont {N.}~\bibnamefont {Qureshi}}, \bibinfo {author}
  {\bibfnamefont {Q.}~\bibnamefont {Faure}}, \bibinfo {author} {\bibfnamefont
  {P.}~\bibnamefont {Bourges}}, \bibinfo {author} {\bibfnamefont
  {V.}~\bibnamefont {Bal\'edent}}, \bibinfo {author} {\bibfnamefont {J.-R.}\
  \bibnamefont {Soh}}, \bibinfo {author} {\bibfnamefont {J.}~\bibnamefont
  {Rau}}, \bibinfo {author} {\bibfnamefont {P.}~\bibnamefont {McClarty}},\ and\
  \bibinfo {author} {\bibfnamefont {A.}~\bibnamefont {Gukasov}},\ }\href@noop
  {} {\bibinfo {title} {Supplemental material
  (url-will-be-inserted-by-publisher)}}\BibitemShut {NoStop}%
\bibitem [{\citenamefont {Buiarelli}\ \emph {et~al.}(2025)\citenamefont
  {Buiarelli}, \citenamefont {Fernandes},\ and\ \citenamefont
  {Birol}}]{Buiarelli2025}%
  \BibitemOpen
  \bibfield  {author} {\bibinfo {author} {\bibfnamefont {L.}~\bibnamefont
  {Buiarelli}}, \bibinfo {author} {\bibfnamefont {R.~M.}\ \bibnamefont
  {Fernandes}},\ and\ \bibinfo {author} {\bibfnamefont {T.}~\bibnamefont
  {Birol}},\ }\bibfield  {title} {\bibinfo {title} {Noncollinear magnetic
  multipoles in collinear altermagnets},\ }\href
  {https://doi.org/10.1103/kq6x-7jfc} {\bibfield  {journal} {\bibinfo
  {journal} {Phys. Rev. B}\ }\textbf {\bibinfo {volume} {112}},\ \bibinfo
  {pages} {224442} (\bibinfo {year} {2025})}\BibitemShut {NoStop}%
\bibitem [{\citenamefont {Becker}\ and\ \citenamefont
  {Coppens}(1974)}]{becker_extinction_1974}%
  \BibitemOpen
  \bibfield  {author} {\bibinfo {author} {\bibfnamefont {P.~J.}\ \bibnamefont
  {Becker}}\ and\ \bibinfo {author} {\bibfnamefont {P.}~\bibnamefont
  {Coppens}},\ }\bibfield  {title} {\bibinfo {title} {Extinction within the
  limit of validity of the {Darwin} transfer equations. {I}. {General}
  formalism for primary and secondary extinction and their applications to
  spherical crystals},\ }\href
  {https://doi.org/https://doi.org/10.1107/S0567739474000337} {\bibfield
  {journal} {\bibinfo  {journal} {Acta Crystallographica Section A}\ }\textbf
  {\bibinfo {volume} {30}},\ \bibinfo {pages} {129} (\bibinfo {year}
  {1974})}\BibitemShut {NoStop}%
\bibitem [{\citenamefont {Delapalme}\ \emph {et~al.}(1978)\citenamefont
  {Delapalme}, \citenamefont {Lander},\ and\ \citenamefont
  {Brown}}]{delapalme_magnetisation_1978}%
  \BibitemOpen
  \bibfield  {author} {\bibinfo {author} {\bibfnamefont {A.}~\bibnamefont
  {Delapalme}}, \bibinfo {author} {\bibfnamefont {G.~H.}\ \bibnamefont
  {Lander}},\ and\ \bibinfo {author} {\bibfnamefont {P.~J.}\ \bibnamefont
  {Brown}},\ }\bibfield  {title} {\bibinfo {title} {Magnetisation density in
  {URh}$_{\textrm{3}}$},\ }\href {https://doi.org/10.1088/0022-3719/11/7/033}
  {\bibfield  {journal} {\bibinfo  {journal} {Journal of Physics C: Solid State
  Physics}\ }\textbf {\bibinfo {volume} {11}},\ \bibinfo {pages} {1441}
  (\bibinfo {year} {1978})}\BibitemShut {NoStop}%
\end{thebibliography}%

\clearpage

\section*{Supplementary Information}
\label{sec:supplementary}

\subsection*{Polarized Neutron Cross Section}

The scattering intensity for polarized neutrons in the elastic channel is given by
\begin{equation}
I^{\pm}(\mathbf{k}) \propto
|N(\mathbf{k})|^2
+ |\mathbf{M}_{\perp}(\mathbf{k})|^2
\;\pm\;
2 \,
\Re\!\Big[
N(\mathbf{k})\,
\mathbf{P}_0 \cdot \mathbf{M}_{\perp}^{\,*}(\mathbf{k})
\Big],
\label{eq:xsection}
\end{equation}
where the nuclear structure factor is
\begin{equation}
N(\mathbf{k}) = 
\sum_{j} b_j\, e^{i\mathbf{k}\cdot\mathbf{r}_j},
\end{equation}
and the magnetic interaction vector is
\begin{equation}
\mathbf{M}_{\perp}(\mathbf{k}) =
\sum_{j}
\frac{r_0 \gamma}{2}\, f_j(\mathbf{k})\, e^{i\mathbf{k}\cdot\mathbf{r}_j}\,
\Big[
\hat{\mathbf{k}} \times 
\big( \mathbf{m}_j \times \hat{\mathbf{k}} \big)
\Big].
\label{eq:mperp}
\end{equation}
Here $\hat{\mathbf{k}}=\mathbf{k}/|\mathbf{k}|$, $f_j(\mathbf{k})$ is the magnetic form factor, and $\mathbf{m}_j$ is the magnetic moment of atom $j$. 
This general expression allows refinement of the magnetic structure using either spherical or multipolar form factors, provided that nuclear and magnetic scattering occur at the same reciprocal-lattice positions, as is the case for MnF$_2$. In Eq.~\ref{eq:xsection} we have omitted the polarization dependent chiral term because it vanishes in MnF$_2$ in the elastic channel as a consequence of the reality of $\mathbf{M}_{\perp}(\mathbf{k})$ which, in turn, comes from the existence of an inversion center.

\subsection*{Data Refinement}

Prior to the polarized‑neutron measurements, the nuclear and magnetic structures
were characterized on the four‑circle diffractometer D9 at the Institut
Laue–Langevin (ILL) using neutrons of wavelength $\lambda = 0.82$~\AA. MnF$_2$
crystallizes in the tetragonal space group $P4_2/mnm$ with lattice parameters
$a=b=4.87$~\AA\ and $c=3.30$~\AA. The Mn ions occupy the $2a$ sites and the
fluorine ions the $4f$ sites. The refined structural parameters,
$x_\mathrm{F}=0.3049(2)$, $B_{\mathrm{iso,F}}=0.388(16)$,
$B_{\mathrm{iso,Mn}}=0.155(35)$, and $\mathrm{Occ}_\mathrm{F}=0.992(5)$, were
used in the subsequent polarized‑neutron analysis.

Polarized‑neutron diffraction refinements for the 2~K, 1~T data set were based on
the measured flipping ratios. The odd reflections, dominated by the spherical
part of the Mn form factor, exhibit extremely large variations in flipping ratio,
ranging from $0.02 < R < 50$. To stabilize the refinement, we used the asymmetry
parameter
\[
\mathrm{As} = \frac{R-1}{R+1},
\]
which balances the weight of reflections with $R>1$ and $R<1$ in the agreement
factor. With an exposure time of 5 minutes per reflection, about 70\% of the
measured asymmetries exceeded the $3\sigma$ threshold for $\mathrm{As}(hkl)$.

In contrast, the flipping ratios of even reflections—sensitive to fluorine
covalence and Mn asphericity—differed from unity by only a few percent. For this
reason, an exposure time of 30 minutes per reflection was required. Because a
complex spin‑density distribution was anticipated, all symmetry‑allowed even
reflections in four unique sets were measured up to
$\sin\theta/\lambda < 0.8$~\AA$^{-1}$. This yielded a 1~T data set consisting of
140 reflections, of which 57 exceeded the $2\sigma$ threshold for
$\mathrm{As}(hkl)$.  The multipolar refinements were performed using the full odd-reflection data set and a threshold-filtered subset of even reflections selected for their high sensitivity to variations in the model parameters. The resulting parameters for 1~T data set are reported in the main text.

\begin{table}[htbp!]
\centering
\caption{Multipolar refinement models for the 2~K, 6~T data set using Slater-type radial functions. Agreement factors are given separately for $\it{all~ odd}$ ($N_\text{peak}=47$) and  $\it{observed~  even}$ reflections ($N_\text{peak}=32$).}
\resizebox{\columnwidth}{!}{
\begin{tabular}{r S S S S}
\hline
 Model & 1 & 2 & 3 & 4 \\
\hline
$m_\text{Mn} P_{00}$ ($\mu_B$) & 5.17(5) & 5.17(5) & 5.17(5) & 5.17(5) \\
$m_\text{F} P_{00}$ ($\mu_B$)   & 0.0      & -0.0275(10) & -0.0301(10) & -0.0277(12) \\
$m_\text{F} P_{20}$ ($\mu_B$)   & 0.0      & 0.0        & -0.0064(8)    & -0.0028(5) \\
$m_\text{Mn} P_{2-2}$ ($\mu_B$) & 0.0      & 0.0        & -0.0167(20)   & 0.0 \\
$m_\text{Mn} P_{4-2}$ ($\mu_B$) & 0.0      & 0.0        & 0.0         & -0.082(8) \\
\hline
$\chi^2/N_\text{peak}$ (odd)  & 5.14 & 5.14 & 5.14 & 5.14 \\
$\chi^2/N_\text{peak}$ (even) & 56.45 & 5.90 & 4.81 & 3.45 \\
\hline
\end{tabular}
}
\label{table:Multi6T}
\end{table}
Refinement of the 6~T data set (Table~\ref{table:Multi6T}), comprising 47 odd and 83 even reflections measured with a 20-minute exposure time, yielded 32 observed even reflections. As for the 1~T data set, the odd reflections are fully described by the spherical Mn magnetic form factor, whereas the even reflections require both a finite ligand spin density and an anisotropic Mn magnetization density. The best agreement is again obtained with Model~4, confirming the presence of the Mn $P_{4-2}$ multipole together with an antiparallel spin polarization on the fluorine ions. As expected upon reversal of the magnetic field, all refined magnetic coefficients change sign while retaining essentially the same magnitudes within the experimental uncertainties.

A refinement of the 6~T data set, with the domain population allowed to vary, shows that following the spin-flop transition, the N$_+$ domain occupies 88(1)\% of the sample volume. The reversal of the Néel vector upon the spin-flop transition reverses the signs of the Mn and F moments, while their magnitudes remain unchanged within the experimental uncertainties. This indicates that the application of a 6~T field has only a minor effect on the magnetic structure.

\begin{figure}[t]
\centering
\includegraphics[width=0.9\columnwidth]{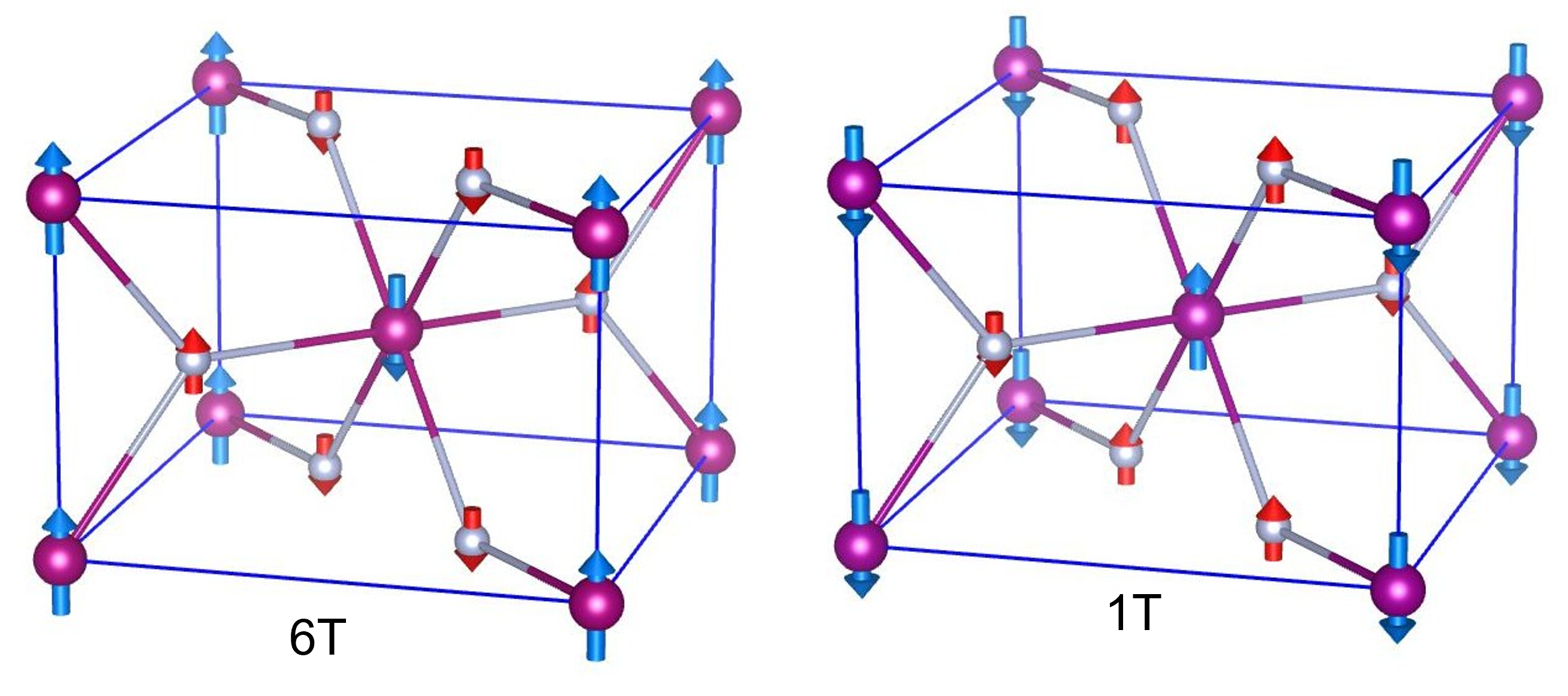}
\caption{Left: magnetic domain $\mathbf{N}_{-}$; right: $\mathbf{N}_{+}$ in MnF$_2$
at 2~K, 1~T and 6~T. Red arrows show Mn moments; blue arrows show fluorine moments
(scaled by $10^2$ for visibility).}
\label{fig:Fig1}
\end{figure}

The corresponding magnetic structures of the refined majority domains, obtained using spherical Model~2 for the 1~T and 6~T data sets, are shown in Fig.~\ref{fig:Fig1}.

Within the experimental precision, no significant field-induced changes in either the ligand polarization or the Mn multipolar coefficients are observed between 1 and 6~T. This robustness indicates that the ferro-octupolar order is governed primarily by the internal Mn molecular field rather than by the externally applied magnetic field.

\subsection*{ Multipolar Modulation of Exchange Interactions}

The refined spin density shows that the Mn magnetization is not purely spherical but contains
a small tesseral hexadecapole $P_{4-2}$. This multipole transforms as
$(x^{2}-y^{2})(7z^{2}-r^{2})$ in the
$(x,y,z)\parallel([110],[1\bar{1}0],[001])$ frame and introduces a four‑lobed anisotropy that
distinguishes the $3d_{xz}$ and $3d_{yz}$ orbitals. This anisotropy directly affects the
Mn–F–Mn superexchange pathways.

For the nearest‑neighbour ferromagnetic path $J_{1}$ (bond angle $101.27^\circ$), both Mn ions
contribute mainly $3d_{xz}$ orbitals. The $P_{4}^{-2}$ multipole enhances the spin density
along $x$, increasing the $3d_{xz}$–$2p_z$–$3d_{xz}$ overlap and strengthening the
ferromagnetic exchange.

\begin{figure}[h]
\centering
\includegraphics[width=0.9\linewidth]{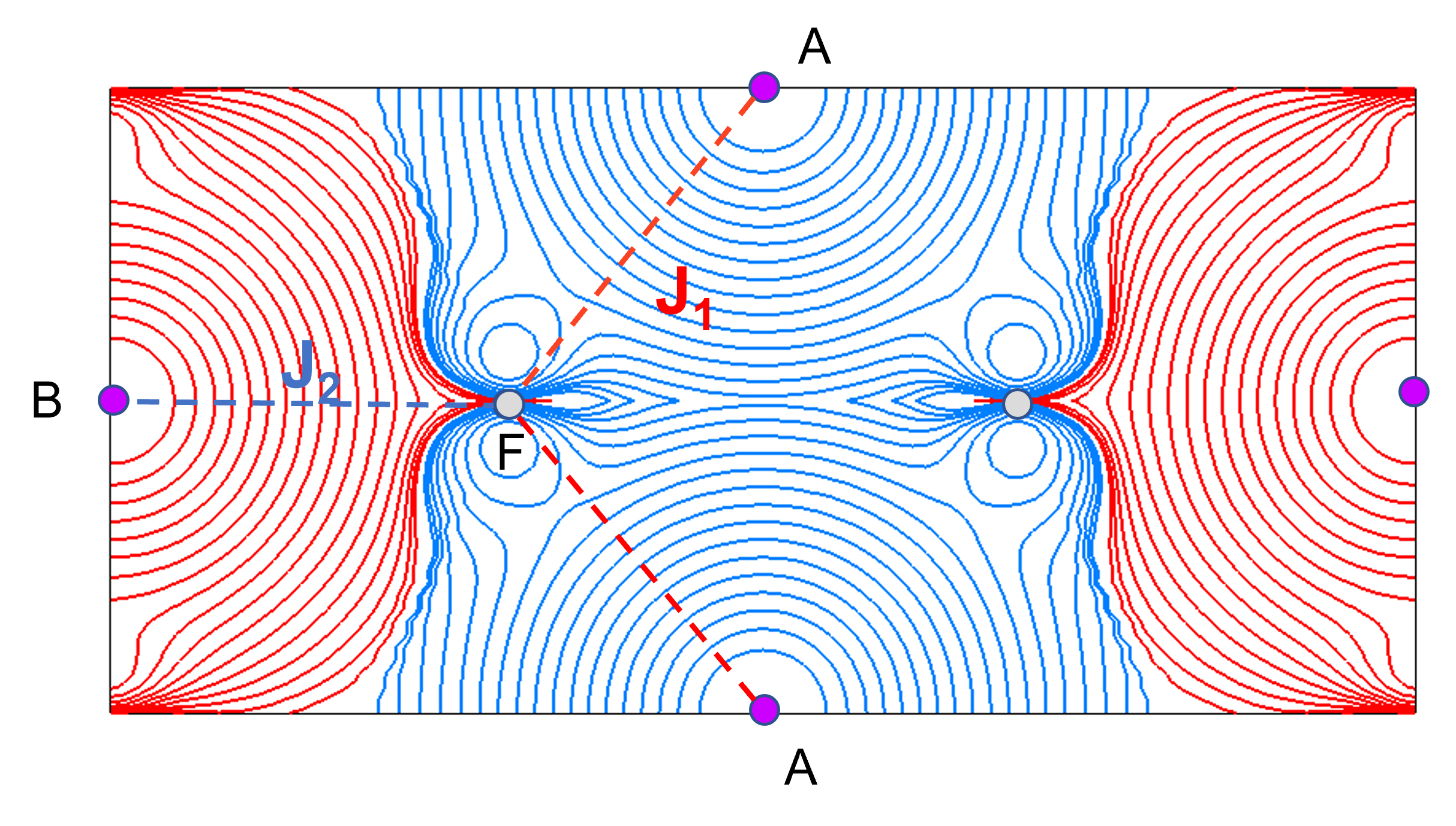}
\caption{
Spin density in the (110) plane centered at the $(\tfrac{1}{2},\tfrac{1}{2},\tfrac{1}{2})$
position, showing the ferromagnetic and antiferromagnetic exchange paths $J_1$ and $J_2$
mediated by the fluorine $p_z$ orbitals.
}
\label{fig:ExchPathJ12}
\end{figure}

The next‑nearest‑neighbour (nnn) path $J_{2}$ ($129.4^\circ$) involves mixed $3d_{xz}$ and $3d_{yz}$
orbitals results in the antiferromagnetic character of
$J_{2}$ in agreement with Goodenough-Kanamori rules.

\begin{figure}[h]
\centering
\includegraphics[width=0.7\linewidth]{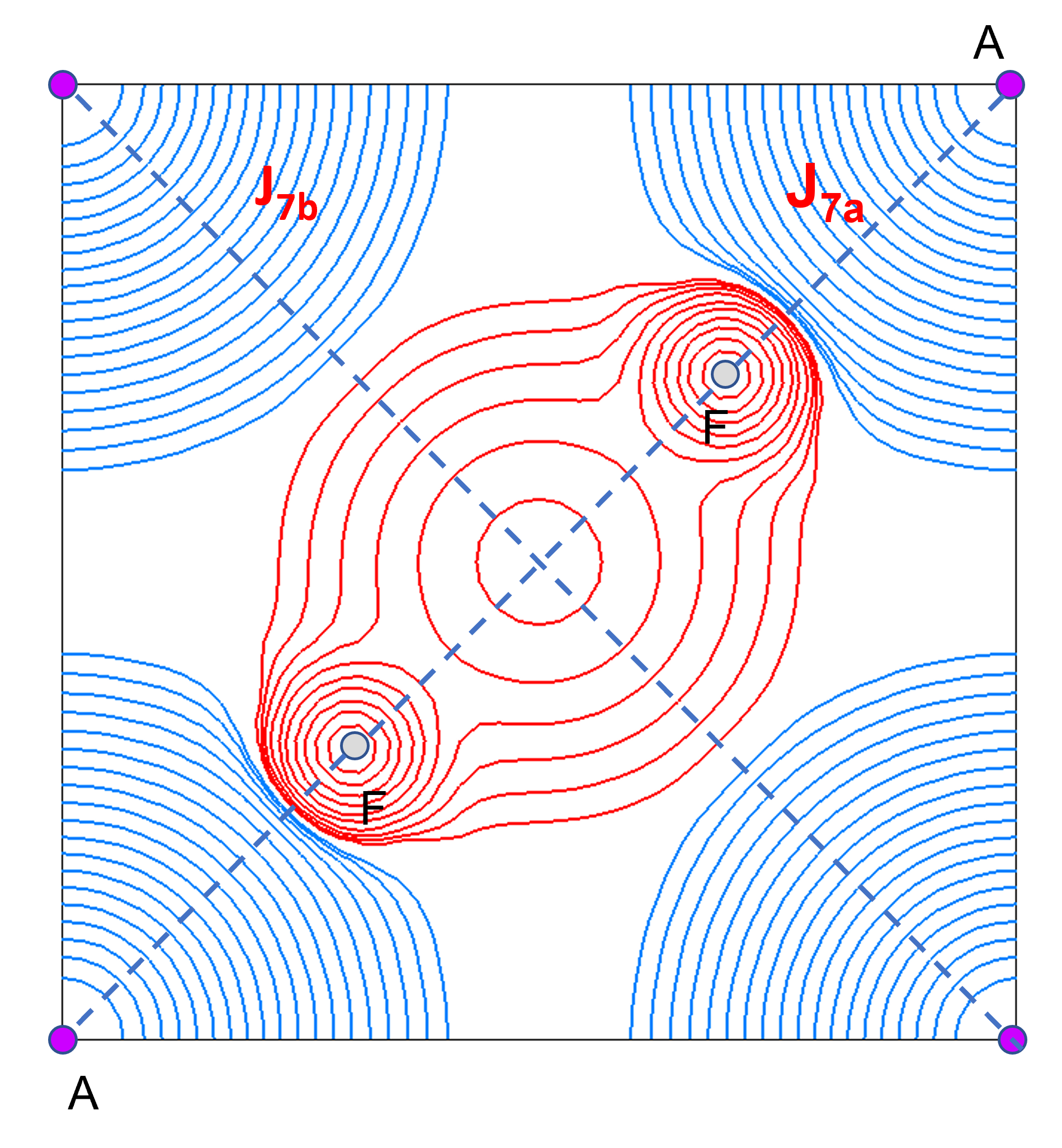}
\caption{
Spin density in the basal plane showing the ferromagnetic exchange path $J_{7a}$ mediated by
hybridization of two fluorine $p_z$ orbitals.
}
\label{fig:ExchPathJ7ab}
\end{figure}

The most significant effect occurs for the $180^\circ$ double‑ligand path $J_{7}$ in the basal
plane. This path couples $3d_{xz}$ orbitals on opposite sublattices through two fluorine
$2p_z$ orbitals. Since $P_{4}^{-2}$ is maximal in the basal plane and changes sign under the
$4_{2}'$ symmetry operation, it induces alternating exchange anisotropies on the two Mn
sublattices. This sublattice‑dependent modulation provides a natural microscopic mechanism for
the emergence of altermagnetic symmetry \cite{Smejkal2023,faure2025}.

Overall, the $P_{4}^{-2}$ hexadecapole is not a minor correction but an active element shaping
the exchange network and reinforcing the altermagnetic topology of MnF$_2$.

\subsection*{Maximum Entropy Method}

To assess the asphericity of the Mn spin density, several MEM reconstructions were
performed on the 1~T and 6~T data sets using deliberately unfavorable, non‑uniform
priors~\cite{papoular}. The prior consists of spherically symmetric magnetization
densities centered on the Mn and F sites, constructed analytically from the refined
moments and the radial integrals obtained in the least‑squares refinement. Because
this prior is strongly biased toward spherical densities, any asphericity emerging
in the MEM solution must be supported directly by the experimental data.

\begin{figure}[h]
\centering
\includegraphics[width=0.8\linewidth]{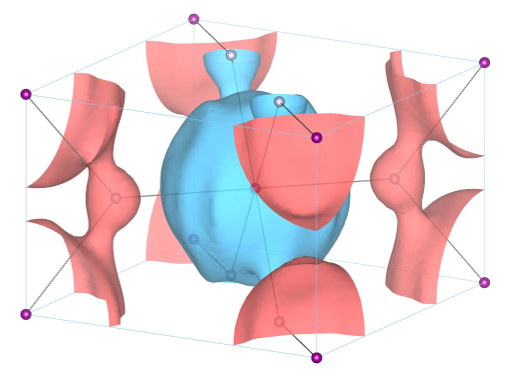}
\caption{
MEM reconstructed spin density   at 2~K in     6~T with  isosurface level taken to be
$0.02\,\mu_B/\text{\AA}^3$.
}
\label{fig:SIFig2}
\end{figure}

Figures~\ref{fig:SIFig2} shows the reconstructed three‑dimensional spin density for
the 6~T data set at 2~K. The MEM reconstruction of the 6~T data containing less statistically significant reflections yields a 
spin‑density distribution very close to that of the 1~T state, apart from
the expected global sign inversion associated with the domain reversal.
Thus, within experimental precision, the application of a 6~T magnetic field does not
produce significant changes in the magnetization density.

\subsection*{Domain Population Control}

Magnetic domain populations were determined prior to each measurement using the polarized--neutron flipping--ratio method of Alperin \textit{et al.}~\cite{Alperin1962}. Bragg intensities for neutron spins parallel ($I^+$) and antiparallel ($I^-$) to the applied field yield the flipping ratio
\[
R = I^+/I^-.
\]

For MnF$_2$, the nuclear and magnetic structure factors of the (320)-type reflections are nearly equal at low temperature, and the ratio reduces to
\[
R = \frac{1 + P_i (2\alpha - 1)}
         {1 - P_i \epsilon (2\alpha - 1)},
\]
where $\alpha$ is the domain fraction and $\epsilon$ the spin--flipper efficiency. This expression enables direct extraction of the domain populations from the measured $R$.

After slow field cooling to $2$\,K in a 5\,T field, flipping ratios of the (3,2,0) and (3,$\bar{2}$,0) reflections were recorded:

At 1\,T, the expected relation
\[
R_{hkl} = 1/R_{h\bar{k}l}
\]
was satisfied, confirming a nearly single antiferromagnetic domain with Néel vector
\[
\mathbf{N}_- = -\mathbf{M}_1(0,0,0),\;
\mathbf{M}_2\!\left(\tfrac{1}{2},\tfrac{1}{2},\tfrac{1}{2}\right),
\]
opposite to the vertical field direction.

MnF$_2$ undergoes a spin--flop transition at $\mu_0 H_{\mathrm{sf}} \approx 9.3$\,T at 2\,K. Due to a slight misalignment of the $c$ axis, the transition occurred at 7.6\,T in our experiment. Reducing the field to 6\,T restored the antiferromagnetic state but with reversed domain populations: refinement shows that the majority domain with Néel vector
\[
\mathbf{N}_+ = \mathbf{M}_1(0,0,0),\;
-\mathbf{M}_2\!\left(\tfrac{1}{2},\tfrac{1}{2},\tfrac{1}{2}\right)
\]
occupies $\sim 88\%$ of the sample, while $\sim 12\%$ remains in the original domain. The 6\,T data set was collected under these conditions.

\begin{table}[h]
\centering
\caption{Measured flipping ratios and extracted domain fractions.}
\begin{tabular}{lccc}
 & Reflection & 2\,K, 1\,T & 2\,K, 6\,T \\
\hline
 & $R_{(3,2,0)}$ & 0.024(4) & 22(2) \\
 & $C_{\rm Dom\,1}$ & $\approx 0\%$ & $\approx 88\%$ \\
 & $R_{(3,\bar{2},0)}$ & 44(4) & 0.042(3) \\
 & $C_{\rm Dom\,2}$ & $\approx 100\%$ & $\approx 12\%$ \\
\end{tabular}
\end{table}

\subsection*{Extinction  Corrections}

In the flipping ratio method, the extinction factors \(y_{+}\) and \(y_{-}\) applied to \(I_{+}\) and \(I_{-}\), respectively are usually obtained from an
extinction model (e.g.\ Becker-Coppens \cite{becker_extinction_1974}) adapted to polarized neutrons. Such an approach is  used in CRYSPY \cite{cryspy}. 

An extinction correction applied to the polarized neutron diffraction has been considered by Delapalme et al. Ref.~\cite{delapalme_magnetisation_1978}. They show that those flipping ratios close to unity are much less sensitive to the extinction as, if one assumes \(y_{+} \approx y_{-}\), the extinction correction cancels in the flipping
ratio. This is the case of \textit{even}  reflections yielding flipping ratio close to 1. 
Therefore in practice, we found that the extinction plays a negligible role  in the refinement and reconstruction of spin density from the \textit{even} data sets, which are related to the asphericity and covalency. 
 
We found that the extinction correction applied to the measured even datasets  did not exceed a few percent in the refined parameters.  However, since the odd reflections,
exhibiting strongly different \(y_{+}\) and \(y_{-}\),  were sensitive to the extinction corrections the 
Becker--Coppens  extinction parameters were applied both to  \textit{odd}  and  \textit{even} datasets. These parameters were obtained by refinement of unpolarized neutron diffraction from D9 diffractometer diffractometer  adapted to polarized neutrons.

\end{document}